\documentclass[12pt]{article}
\pdfoutput=1  

\usepackage{amssymb}
\usepackage{amsmath}
\usepackage{graphicx}
\usepackage{caption}
\usepackage{subcaption}
\usepackage[figuresright]{rotating}
\usepackage[]{algorithm2e}
\usepackage{soul}
\usepackage{url}
\usepackage[colorinlistoftodos]{todonotes}
\usepackage{authblk}
\usepackage{soul}

\newcommand{\R}[1]{\mathbb{R}^{#1}}
\newcommand{\twod}{\ensuremath{\ell_2^2}}
\newcommand{\threed}{\ensuremath{\ell_2^3}}
\newcommand{\oT}{LRT} 
\newcommand{\ourTree}{Linear Regression Tree} 

\newcommand{\h}[1]{{#1}}

\begin{document}

\title{Supermetric Search} 

\author[1]{Richard Connor}
\affil[1]{Department of Computer and Information Sciences,University of Strathclyde, Glasgow, G1 1XH, United Kingdom}
\author[2]{Lucia Vadicamo}
\author[3]{Franco Alberto Cardillo}
\author[2]{Fausto Rabitti}
\affil[2]{Institute of Information Science and Technologies (ISTI), CNR, Pisa, Italy}
\affil[3]{Institute of Computational Linguistics (ILC), CNR, Pisa, Italy}
\affil[1]{{\footnotesize richard.connor@strath.ac.uk}}
\affil[2]{\footnotesize  \{lucia.vadicamo, fausto.rabitti\}@isti.cnr.it}
\affil[3]{\footnotesize  francoalberto.cardillo@ilc.cnr.it}

\date{}                     
\renewcommand\Affilfont{\itshape\small}
\maketitle

\begin{abstract}
  Metric search  is  concerned with the efficient evaluation of queries in metric spaces. In general, a  large space of objects is arranged in such a way that, when a further object is presented as a query, those objects most similar to the query can be efficiently found. Most  mechanisms rely upon the triangle inequality property of the metric governing the space. The triangle inequality property  is equivalent to a finite  embedding property, which states that any three points of the space can be isometrically embedded in two-dimensional Euclidean space. In this paper, we examine a class of semimetric space which is finitely four-embeddable in three-dimensional Euclidean space. In mathematics this property has been  extensively studied and is generally known as the \emph{four-point} property. 
 %
 All spaces with the four-point property are metric spaces, but they also have some stronger geometric guarantees. We coin the term \emph{supermetric%
 	\footnote{This term has previously been used in the domains of particle physics and evolutionary biology as a pseudonym for the mathematical term \emph{ultra-metric}, a concept of no interest in metric search; we believe our concept is of sufficient importance to the domain to justify its reuse with a different meaning.}
 	space} as, in terms of metric search, they are significantly more tractable. Supermetric spaces include all those governed by Euclidean, Cosine\footnote{for the correct formulation of Cosine distance, see \cite{hilbert_embedding} for details}, Jensen-Shannon and Triangular distances, and are thus commonly used within many domains. In previous work we have given a generic mathematical basis for the supermetric property and shown how it can improve indexing performance for a given exact search structure. Here we present a full investigation into its use within a variety of different hyperplane partition indexing structures, and go on to show some more of its flexibility by examining a search structure whose partition and exclusion conditions are tailored, at each node, to suit the individual reference points and data set present there. Among the results given, we show a new  best performance for exact search using a well-known benchmark.

\end{abstract}

%
%
%

\renewcommand{\arraystretch}{1.1}
\section{Introduction}

 Within any metric space, any three objects can be used to construct a triangle in 2D Euclidean space, where the objects are represented by the vertices of the triangle and the edges preserve their distances in the original space. That is, any metric space is isometrically three-embeddable in 2D Euclidean space.
 
 Some metric spaces are also isometrically four-embeddable in 3D Euclidean space, allowing the construction of a tetrahedron.
 We  have  previously shown how these spaces have  further geometric properties which can be used to improve the performance of exact search,  in particular for any search mechanism based on hyperplane partitioning.  This leads to the notion of a \emph{supermetric space} \cite{Connor2016:supermetricSISAP},  a space which is also a metric space but with further geometric properties which give stronger guarantees for search mechanisms.
 Furthermore, we have given a rigorous and constructive mathematical basis for assessing whether a proper metric  space has the supermetric property, and showed how this property allows the use of the Hilbert Exclusion mechanism in place of the less powerful hyperbolic exclusion \cite{hilbert_embedding}. In \cite{Connor2016:supermetricSISAP}, we also showed how the supermetric property could, in principle, be used to construct arbitrary partitions within a 2D plane into which many objects are projected, due to a lower-bound property which is a  corollary of the four-point property.
 
 {In this paper we extend initial work which appeared in} \cite{Connor2016:supermetricSISAP}{ by taking the investigation to its next stage.}
 While we previously showed how the use of the Hilbert Exclusion property gave a significant improvement in performance when used in conjunction with  a particular state-of-the-art hyperplane-based indexing mechanism, the Distal Spatial Approximation Tree (DiSAT, \cite{dSatIS}), we now perform a full evaluation over its performance within a fully general context of twelve different hyperplane tree indexing structures. The outcome is that a simpler data structure is found to be the most efficient in this context, and indeed gives a new best-published performance for threshold search over the SISAP benchmark data sets \cite{SISAP_man}; to put this result in perspective, it requires only around 40\% of the number of distance calculations per query of the previous state of the art given in \cite{dSatIS}.
 
 Beyond these benchmark data sets, which in this context are relatively small and tractable, we show the performance advantages hold in some larger data sets, as dimensionality and object size increase, and also for a number of different distance metrics.
 
 Further, we begin to investigate more flexible use of the planar lower bound property we first described in \cite{Connor2016:supermetricSISAP}. At the time of this publication we observed that the property was more general than the Hilbert Exclusion property. Now we are able to show a remarkably flexible use within a   hyperplane tree built over ``real-world" data sets; the significance of such data is that it is typically distributed in a non-uniform manner within the space. 
 The non-uniformity manifests differently with each choice of reference points, and this data structure allows a different strategy to be used in each node, to maximise the advantage which can be gained.

 The rest of this paper is organised as follows.
 Section \ref{sec:RW} sets the detailed technical context for the work, including related work by ourselves and others. 
 Section \ref{sec:Proj}  then explains a novel observation which is a consequence of the four-point property: tetrahedral projection onto a plane, which gives an important lower-bound property. In fact, it turns out that Hilbert Exclusion results as a simple corollary of this more general property. In this section we discuss a number of relatively deep results which are consequent to the property.
 Section \ref{sec_linear_regression} then fully defines a completely novel indexing structure which is only possible to use in a supermetric space, where the hyperplane partition and consequent exclusion mechanism are dynamically chosen according to the distribution of data within each individual node of the tree. 
 
 Section \ref{sec:exp} takes as its starting point the observation that the best indexing techniques are likely to be different in the supermetric context, and gives a full investigation of various hyperplane trees in order to determine the most suitable. Section \ref{section_hilbert_costs}  analyses the extra cost required to make use of the supermetric properties for hyperplane indexing, which in fact is very small. Section \ref{section_large_experiments}  examines the use of the best data structures identified through their application to a number of large and real-world data sets, in order to test their performance in more general contexts.
 
 Finally in Section \ref{sec:conclusion} we give some conclusions and outline areas of further work.

\section{Preliminaries and Related Work}\label{sec:RW}
\renewcommand{\arraystretch}{1.3}
\begin{table}[tb]
	{\small
		\begin{tabular}{|p{0.15\columnwidth} | p{0.8\columnwidth}|}
			\hline
			\textbf{Notion} & \textbf{Definition}  \\	\hline\hline
			$(U,d)$ &  the data domain $U$ and the metric distance $d:U \times U \to \R{}$ \\ \hline
			$S$ & finite set of data objects, $S \subseteq U$\\\hline
			$x,y,s$	&  generic data object, $x,y,s\in S$ \\ \hline
			$q$		& query object $q\in U$  \\ \hline
			$t, t_i$	& threshold distances used in the range search, $t, t_i\in \R{}$  \\ \hline
			$R$ &   solution set for a range query: $R=\{s \in S |\, d(q,s) \le t\}$ \\ \hline
			$\{p_1,\dots,p_n\}$ & set of  $n$ pivots, $p_i \in U$  \\ \hline
			$\ell_2$ & Euclidean \textit{distance}: $\ell_2 (x,y)=\sqrt{\sum_{i=1}^n(x_i-y_i)^2}$ for $x,y\in \R{n}$ \\ \hline
			$\ell_2^n$ & n-dimensional Euclidean \textit{space}, \textit{i.e.} $(\R{n}, \ell_2)$. \\ 
			& For example, $\ell_2^n=(\R{n}, \ell_2)$, $\ell_2^3=(\R{3}, \ell_2)$\\\hline
			Isometric 3-embedding in $\twod$ & A metric space $(U,d)$ is isometrically $3$-embeddable in {\twod} if for any three points $x_1,x_2, x_3 \in U$ there exists a function $f:U \to \ell_2^2$ such that $\ell_2(f(x_i), f(x_j))=d(x_i,x_j)$, for $i,j=1,2,3$\\\hline
			Isometric 4-embedding in $\ell_2^3$ & A metric space $(U,d)$ is isometrically $4$-embeddable in $\ell_2^3$ if for any four points $x_1,x_2, x_3, x_4 \in U$ there exists a function $f:U \to \ell_2^2$ such that $\ell_2(f(x_i), f(x_j))=d(x_i,x_j)$, for $i,j=1,2,3,4$\\ \hline
			four-point property & A metric space $(U,d)$  has the four-point property if it is  isometrically 4-embeddable in $\ell_2^3$\\ \hline
			$\overline{vw}$ & line between two points $v,w\in \R{n}$\\
			\hline
	\end{tabular} }
	\caption{Notation used throughout this paper}	\label{tab:notation} 
\end{table}
To set the context, we are interested in searching a (large) finite  set of objects $S$ which is a subset of an infinite set $U$, where $(U, d)$ is a \textit{metric space}. 
A metric space is an ordered pair $(U, d)$, where $U$ is a domain of objects and $d$ is a total distance function $d:U \times U \to \R{}$, satisfying postulates of non-negativity, identity, symmetry, and triangle inequality \cite{zezula2006similarity}.
The general requirement is to efficiently find members of $S$ which are similar to an arbitrary member of $U$, where the distance function $d$ gives the only way by which any two objects may be compared - the bigger the distance $d(x,y)$, the less similar the data objects $x,y\in U$.
There are many important practical examples captured by this mathematical framework, see for example \cite{Chavez05,zezula2006similarity}.  
Such spaces are typically searched with reference to a query object $q \in U$.
The simplest type of similarity query is the \textit{range search} query. 
A {range search}  for some threshold $t$, based on a query $q \in U$, has the solution set  $R=\{s \in S |\, d(q,s) \le t\}$. 
Other forms of search, for example \textit{nearest neighbor search} (i.e. find the $k$ closest objects to a query), are also useful; here we are studying mostly properties of spaces in general and restrict our attention to the scenario outlined.

Symbols and abbreviations used throughout this paper are summarized in Table \ref{tab:notation}.
\subsection{Metric indexing}
Typically, the distance function is too expensive or $S$ is too large to allow an exhaustive search, that is a sequential scan of the entire dataset. The retrieval process is facilitated by using a \textit{metric index}, one of a large family of data structures used to preprocess the data in such a way as to minimise the time required to retrieve the query result. This data structure can be expensive to build, but this cost is amortized by saving I/O and distance evaluations over several queries to the database. In general, the triangle inequality property is exploited to determine subsets of $S$ which do not need to be exhaustively checked.  Such avoidance is normally referred to as \emph{exclusion} or \emph{space pruning}.

For exact metric search, almost all indexing methods can be divided into those which at each exclusion possibility use  a single ``pivot" point to give \textit{radius-based exclusion}, and those which use two reference points to give \textit{hyperplane-based exclusion}. Many variants of each have been proposed, including many hybrids;~\cite{zezula2006similarity},~\cite{Chavez:2001}, and \cite{Hjaltason2003}  give excellent surveys.
In general the best choice seems to depend on the particular context of metric and data.

Here our focus is particularly on  mechanisms which use hyperplane-based exclusion. 
The simplest such index structure is the Generalised Hyperplane Tree (GHT)~\cite{GHT}. Others include Bisector trees~\cite{Kalantari1983} and variants on them (\textit{e.g.}  Monotonous Bisector Trees~\cite{Noltemeier1992} and Voronoi Trees~\cite{Dehne1987}), the Metric Index \cite{MIndex2011}, and the Spatial Approximation Tree~\cite{SAT2002}. This last has various  derivatives, notably including the Dynamic SAT~\cite{DSAT} and the Distal SAT (DiSAT)~\cite{dSatIS}.

\subsection{Metric Spaces and Finite Isometric Embeddings}
\begin{figure}[tbp]
	\begin{center}
		{\includegraphics[trim=30mm 20mm 30mm 30mm, width=0.5\textwidth]{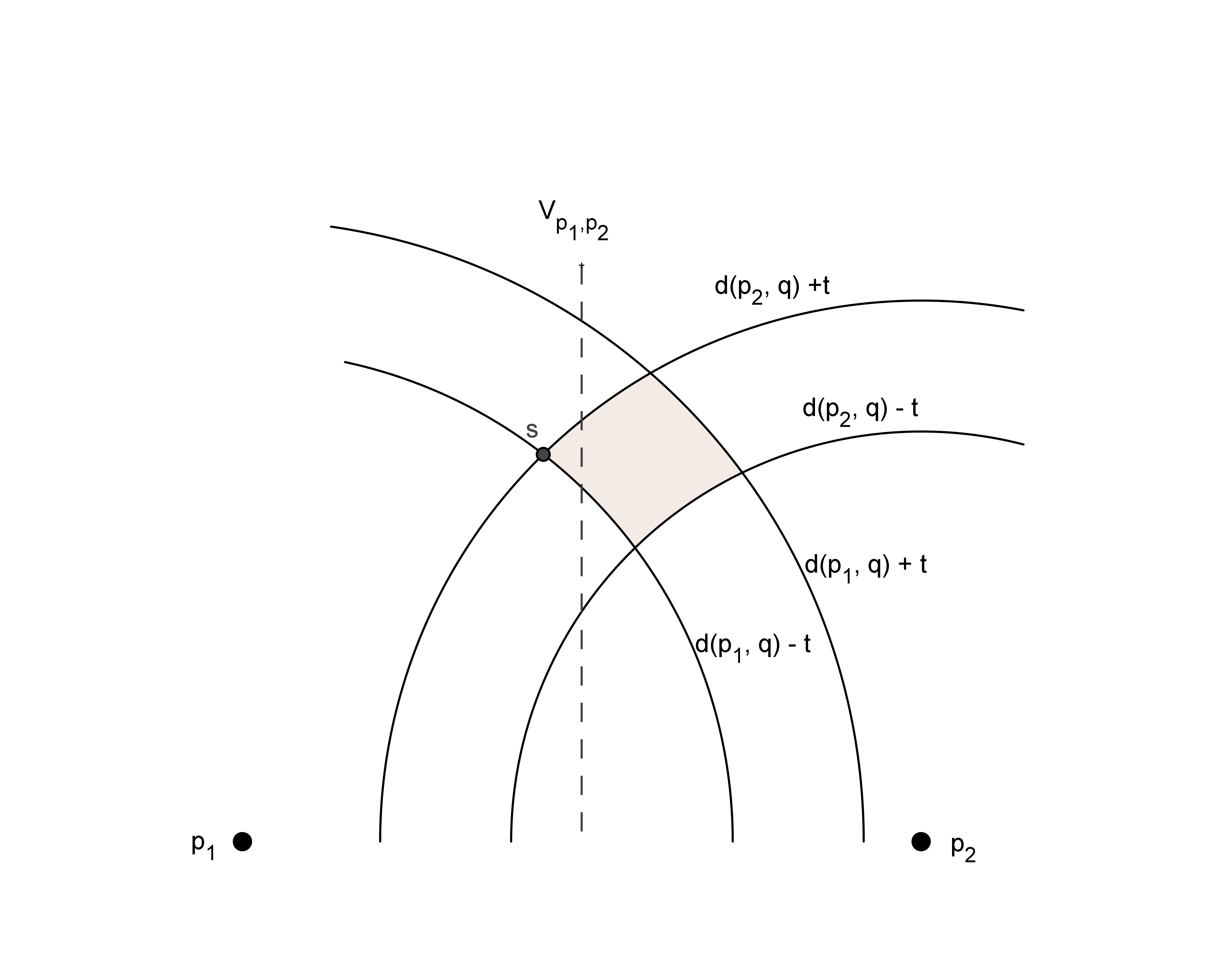}}%
		\caption{In any metric space,  two pivot points $p_1, p_2$ and any solution to a query $q$ can be isometrically  embedded in \twod.  The point $q$   cannot  be drawn in the same diagram. Given its distance from $p_1$ and $p_2$,  any solution in the original metric space must lie in the region bounded by  the four arcs shown. If the  point $s$ lies to the right of $V_{p_1,p_2}=\{x\in S|\, d(x,p_1)=d(x,p_2)\} $,  there is therefore no requirement to search to the left of the hyperplane in the original space. In general, when relying only on the triangle inequality, then half of the space can be excluded from the search only if $|d(q,p_1)-d(q,p_2)|>2t$.
		}
		\label{figure_hyperplane_principle}
	\end{center}
\end{figure}
An  isometric embedding of one metric space $(V,d_v)$ in another $(W,d_w)$ can be achieved when there exists a mapping function $f:V\rightarrow W$ such that $d_v(x,y) = d_w(f(x),f(y))$, for all $x,y \in V$. A finite isometric embedding occurs whenever this property is true for any finite selection of $n$ points from $V$, in which case the terminology used is that $V$ is isometrically $n$-embeddable in $W$.

The idea of characterising a space metrically by means of ``\textit{n-point relations}" seems to have originated in the paper \cite{detilly1892} published in 1892 by de Tilly, a Belgian artillery officer. Some of the questions raised by de Tilly were answered by some mathematicians of the late 19th century, and only in 1928 Karl Menger~\cite{Menger1928} provided a first systematic development of abstract \textit{distance geometry}. The interest of the distance geometry is in all those of transformations of sets for which the distance of two points is an invariant. 
So, as highlighted by Blumenthal~\cite{blumenthal1953}, ``\textit{distance geometry may operate in any kind of space in which a notion of ``distance'' is attached to any point-pair of the space}''.

\paragraph{Isometric 3-embedding in \twod}
The first observation to be made in this context is that any metric space {$(U,d)$} {is \emph{isometrically $3$-embeddable in \twod}, i.e. for any three points $x_1,x_2,x_3 \in U$ there exists a mapping function $f:(U,d)\to (\R{2}, \ell_2)$ such that $\ell_2\left(f(x_i),f(x_j)\right)=d(x_i,x_j)$, for $i,j=1,2,3$. This is  apparent from the triangle inequality property of a proper metric. In fact the two properties are  equivalent: for any semi-metric space 
	which is isometrically $3$-embeddable in  \twod, triangle inequality also holds.  It is interesting to consider the standard exclusion mechanisms of pivot-based exclusion and hyperplane-based exclusion in the light of an isometric 3-embedding in \twod; Figure \ref{figure_hyperplane_principle} for example shows a basis for hyperplane exclusion using only this property rather than triangle inequality explicitly. 
	\paragraph{Supermetric Spaces: Isometric 4-embedding in \threed}
	%
	%
	
	It turns out that many useful metric spaces have a stronger property: they are \emph{ isometrically 4-embeddable in \threed}, which means that for any four points in the space there exists an embedding into $(\R{3},\ell_2)$ that preserves all the ${4 \choose 2} =6$ interpoint distances. In the mathematical literature, this has been referred to as the \emph{four-point property}~\cite{blumenthal1953}. Wilson~\cite{wilson1932relation} shows various properties of such spaces, and Blumenthal~\cite{blumenthal1953} points out that results given by Wilson, when combined with work by Menger~\cite{Menger1928}, generalise to show that some spaces have the \textit{n-point property}: that is, any $n$ points can be isometrically embedded in a Euclidean  $(n-1)$-dimensional space.

	We have studied such spaces in the context of metric indexing in \cite{hilbert_embedding}, where we develop in detail the following   outcomes:
	\begin{enumerate}
		\item Any metric space which is isometrically embeddable in a Hilbert space\footnote{
			A Hilbert space $H$ can be thought of as a generalization of Euclidean space to any finite or infinite number of dimensions. It is an inner vector space which is also a complete metric space with respect to the distance function induced by the inner product. This means that it has an inner product $<\cdot,\cdot>:H\times H \to \mathbb{C}$ that induces a distance $d(\cdot,\cdot)=\sqrt{<\cdot,\cdot>}$ such that every Cauchy sequence in $(H,d)$ converges to a point in $H$ (intuitively, there are no ``points missing'' from $H$).
		}
		has the {\textit{n-point} property}, and so the \textit{four-point} property as well.  
		\item Important spaces with the \textit{n-point} property include, for any dimension, spaces with the following metrics: Euclidean, Jensen-Shannon, Triangular, and (a variant of) Cosine distances.
		\item Important spaces which do not have the \textit{four-point} property include those with the metrics: Manhattan, Chebyshev, and Levenshtein distances.
		\item However, for any metric space $(U,d)$, the space $(U,{d}^\alpha)$, $0<\alpha\leq\frac{1}{2}$ does have the \textit{four-point} property.
	\end{enumerate}
	%
	%
	
	In terms of practical impact on metric search, in \cite{hilbert_embedding} we show only how the four-point property can be used to improve standard hyperplane partitioning. We consider a situation where a subspace is divided according to which of two selected reference points $p_1$ and $p_2$ is the closer. When relying only on triangle inequality, that is in a metric space without the four-point property, then for a query $q$ and a query threshold $t$, the subspace associated with $p_1$ can be excluded from the search only if $d(q,p_1) - d(q,p_2) > 2t$. As the region defined by this condition when projected onto the plane is a hyperbola (see Figure \ref{figure_hyperplane_principle}), we name this \textit{Hyperbolic Exclusion}\footnote{In the literature, the Hyperbolic Exclusion is also referred to as \textit{Double-Pivot Distance Constraint}~\cite{zezula2006similarity}.}.
	
	If the space in question has the \textit{four-point} property, however, we  show that, for the same subspaces, there is no requirement to search that associated with $p_1$ whenever {$$\frac{d(q,p_1)^2  - d(q,p_2)^2}{d(p_1,p_2)} > 2t;$$} this is a weaker condition and therefore allows, in general, more exclusion. We name this condition \textit{Hilbert Exclusion}.
	
	A formula equivalent to  {$$\frac{|d(q,p_1)^2  - d(q,p_2)^2|}{2d(p_1,p_2)}$$} has been used in the context of metric search also in~\cite{Pola2014,Mic2015}, in order to \emph{estimate} the distance between the point $q$ and the hyperplane equidistant from $p_1$ and $p_2$. This formula was derived using the cosine law and was applied only with distances on metric space with the ``semidefinite positive property"~\cite{Pola2014,Schoenberg1938}, since this property allows defining a notion of ``angle" in a generic metric space. To provide a bridge to our work, we observe that the semidefinite positive property is equivalent to the \textit{n-point} property for a finite semimetric space (see Chapter IV, Section 43 of \cite{blumenthal1953}).
	
	In this paper, we examine a more general consequence of  four-point embeddable spaces and show some interim results including new best-performance search of  SISAP data sets. 
	
	%
	%

\section{Tetrahedral Projection onto a Plane} \label{sec:Proj}
In a supermetric space, any two reference points $p_1$ and $p_2$, and query point $q$, \emph{and} any solution to that query $s$ where $d(q,s) \le t$, can \emph{all} be embedded in 3D Euclidean space. As such, they can be used to form the vertices of a tetrahedron. It seems that, while simple metric search is based around the  properties of a triangle, there should be corresponding tetrahedral properties which give a new, stronger, set of guarantees.

Assume that for some search context, points $p_1, p_2 \in U$ are somehow selected and a data structure is built  for a finite set $S \subset U$ where, for $s \in S$, the three distances $d(p_1,p_2), d(s,p_1)$ and $d(s,p_2)$ are calculated during the build process and used to guide the structuring of the data. At query time, for a query $q$, the two distances $d(q,p_1)$ and $d(q,p_2)$ are calculated and may be used to make some deduction relating to this structure.

This situation gives  knowledge of two adjacent faces of the tetrahedron which can be formed in three dimensions. Five of the six edge lengths have been measured, and the final edge is upper-bounded by the value of $t$. Therefore, for a point $s$ to be a solution to the query, it must be possible to form a tetrahedron with the five measured edge lengths, and a last edge of length $t$.
\begin{figure}[t]
	\centering
	{\includegraphics[trim=0mm 5mm 5mm 0mm, width=0.45\textwidth]{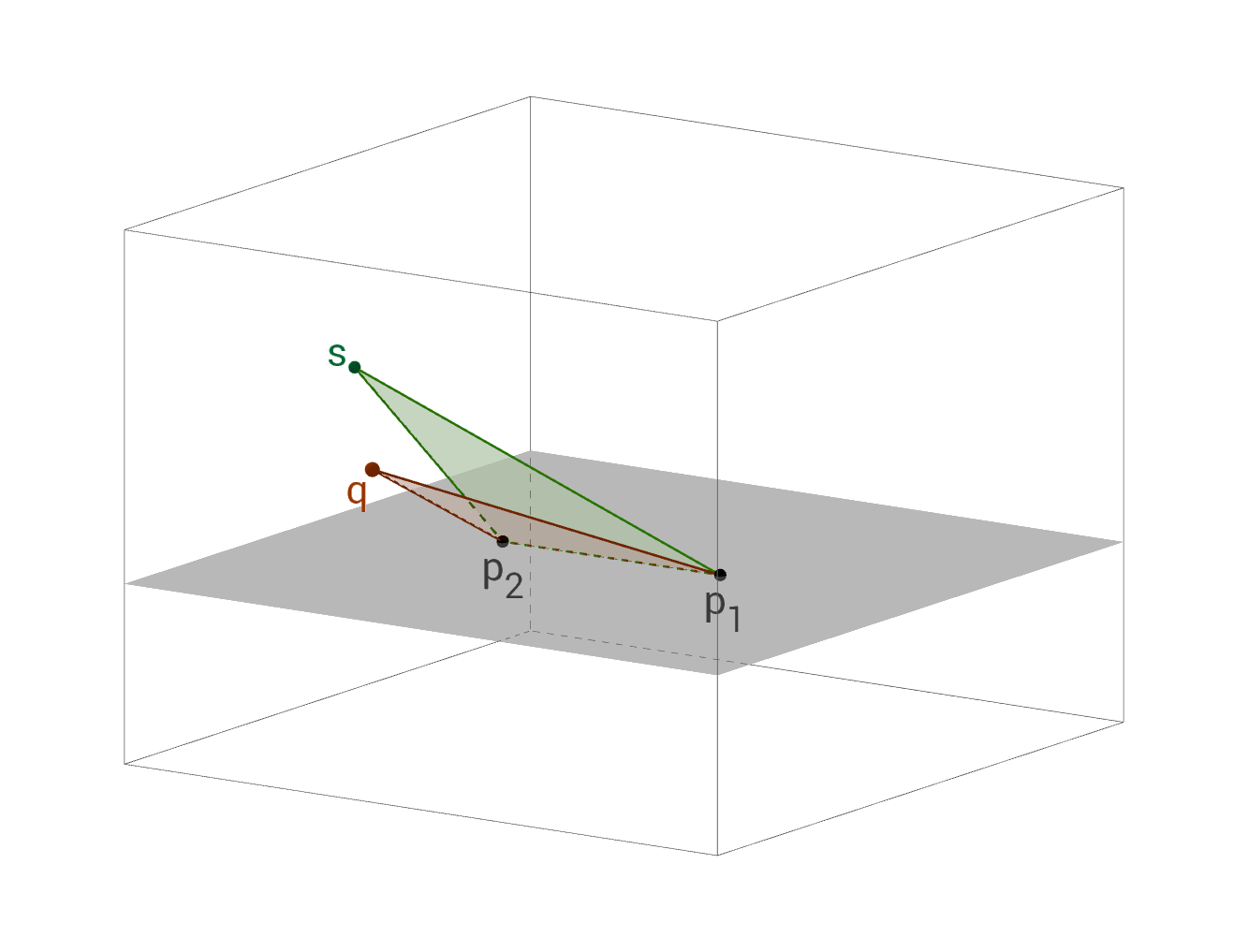}}
	\caption{Two triangles with a common base in 3D space}%
	\label{fig_two_triangs}
\end{figure}
%

%
Figure \ref{fig_two_triangs} shows a situation where five edge lengths have been embedded in 3D space. The edge {$\overline{p_1p_2}$} is shared between the two facial triangles depicted.  However the distance $d(s,q)$ is not known, and therefore neither is the angle between these triangles. The   observation which gives rise to the results presented here is that, if both triangles are now projected onto the same plane, which can be achieved  by rotating one of them around the line {$\overline{p_1p_2}$} until it is coplanar with the other, then for any case where the final edge of the  tetrahedron ($qs$) is less than the length $t$, then the length of this side in the resulting planar tetrahedron is  upper bounded by $t$, as illustrated in Figure \ref{fig_triangsCoplanar}.
\begin{figure}[bt]
	\centering
	{	\includegraphics[trim=15mm 5mm 10mm 20mm, width=0.45\textwidth]{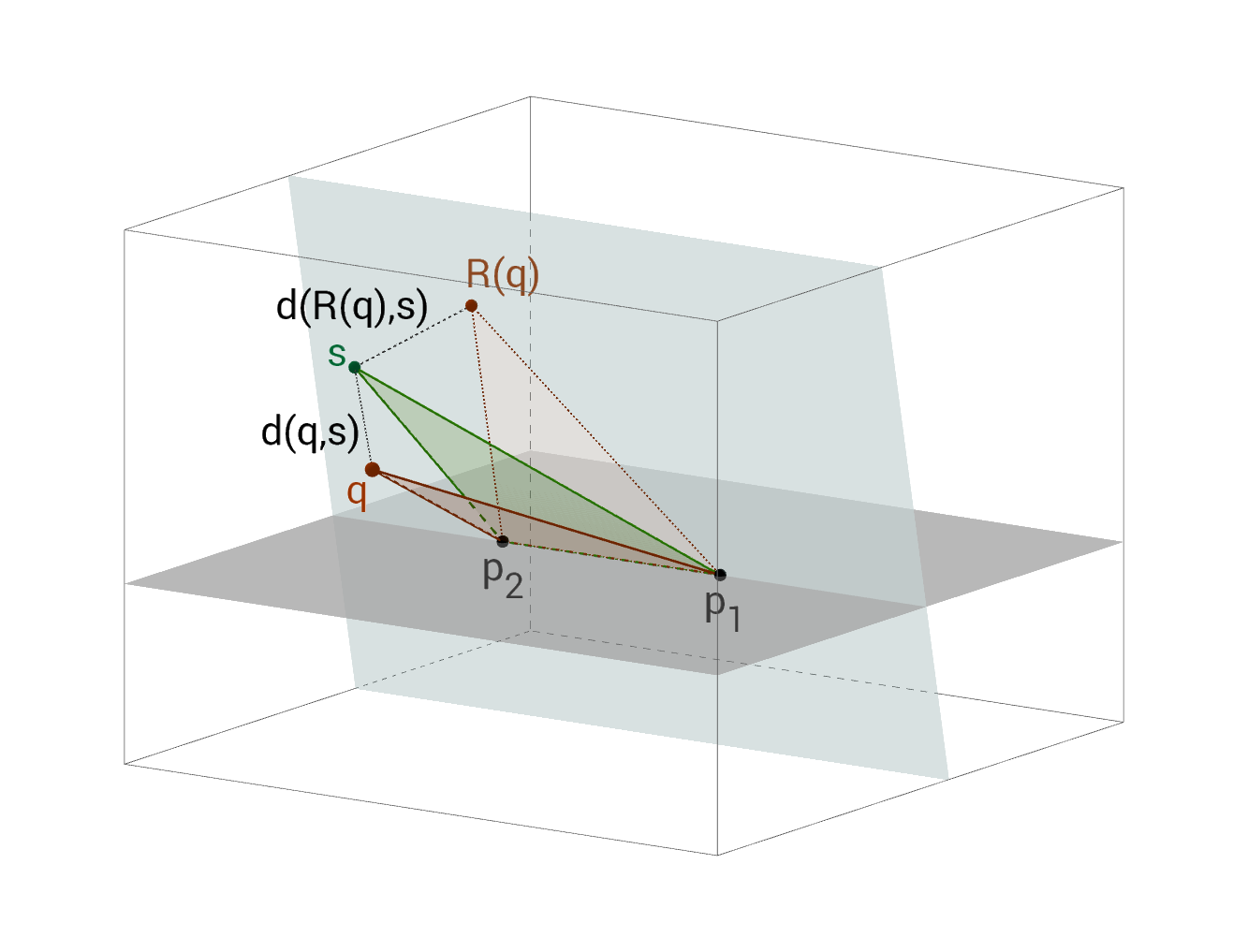}}
	{	\includegraphics[trim=40mm 10mm 20mm 5mm, width=0.44\textwidth]{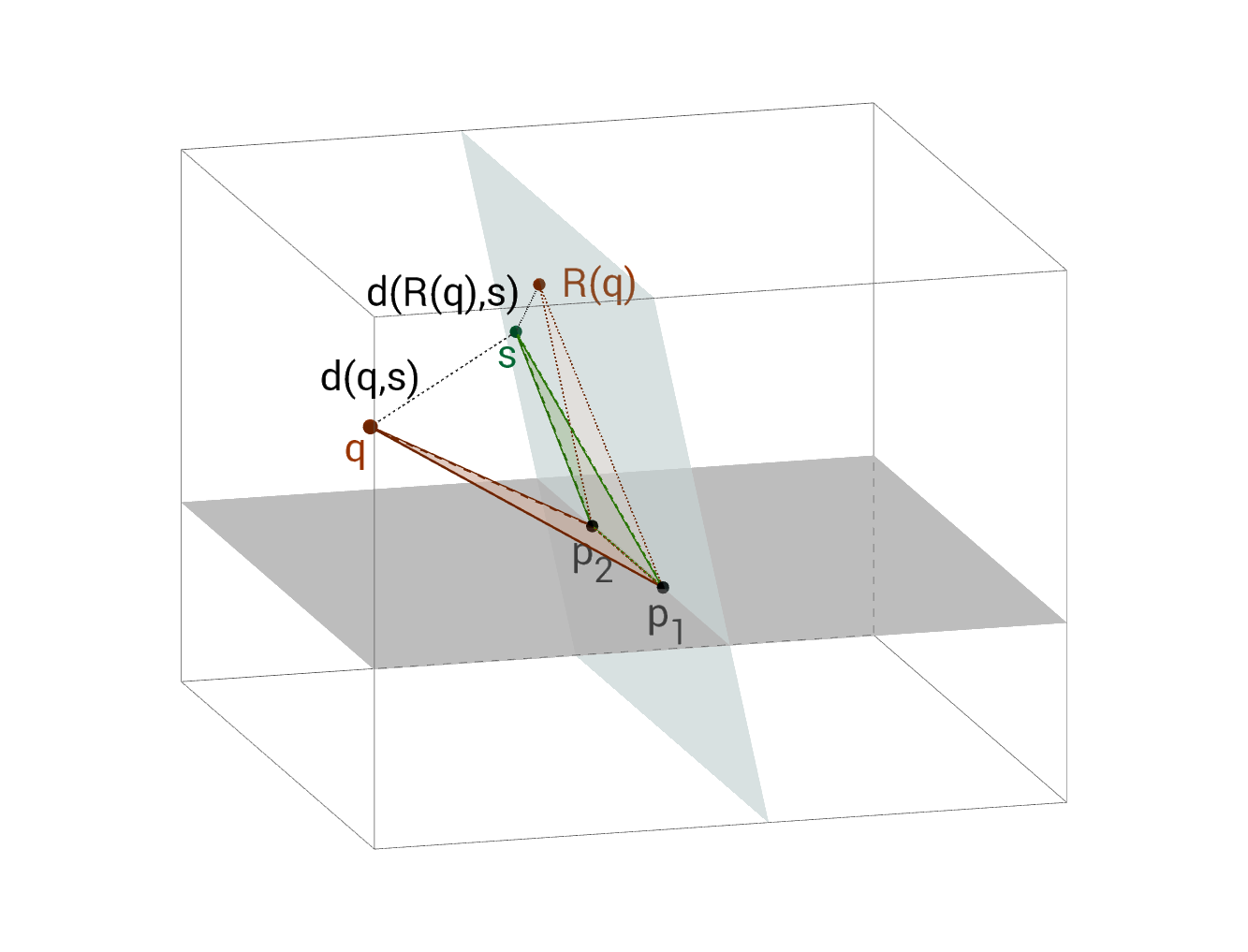}}
	\caption{Projection of the two triangles onto the same plane  by rotation around  {$\overline{p_1p_2}$}. Note that $\ell_2^2(R(q),s) \le \ell_2^3(q,s)${, where R(q) is the point obtained by rotating $q$ around the line $\overline{p_1p_2}$ until it is coplanar with $s$}}
	\label{fig_triangsCoplanar}
\end{figure}

%
Many such coplanar triangles can be depicted, representing many points in a single space, in a single scatter plot as in Figure \ref{fig_simple_scatter}. This shows a set of 500  points, drawn from randomly generated 8-dimensional Euclidean space, and plotted with respect to their distances from two fixed reference points $p_1$ and $p_2$. The distance between the reference points is measured, and the reference points are plotted on the X-axis symmetrically either side of the origin. For each point in the rest of the set, the distances $d(s,p_1)$ and $d(s,p_2)$ are calculated, and used to plot the unique corresponding point in a triangle above the X-axis, according to these edge lengths. In this figure, in consideration with our observations over Figure 
\ref{fig_triangsCoplanar}, it can be seen that, if any two  points  are separated by less that some constant $t$ in the original space, and thus also in the 3D embedding, then they are also within $t$ of each other in this scatter plot.

\begin{figure}[tb]
	\centering
	\includegraphics[trim=50mm 50mm 40mm 0mm, width=0.5\textwidth]{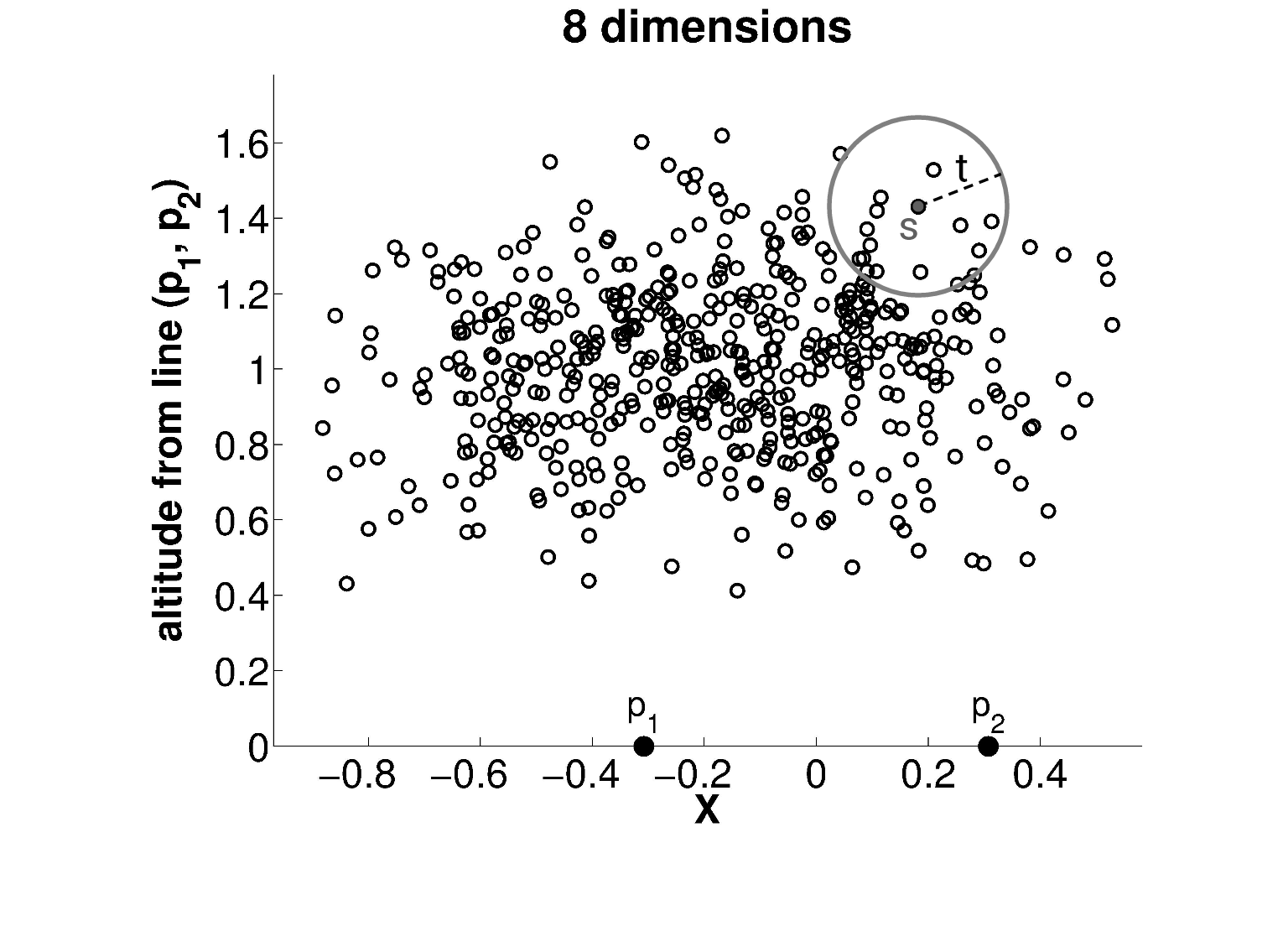} 
	\caption{Scatter diagram for 8-dimensional Euclidean Space. The distance $\delta$ between two selected reference points  $p_1$ and $p_2$ is measured, and an embedding function is chosen which maps these to 
		$(0,-\delta/2)$ and $(0,\delta/2)$ respectively. Other points $s_i$ in the space are then plotted to preserve the distances $d(s_i,p_1)$ and $d(s_i,p_2)$. For metric spaces with the four-point property, the $\ell_2$ distance between the corresponding points in this diagram is a lower bound on  $d(s_i,s_j)$ in the original space. 
		Hence, any point within $t$ of a point $s$ in the original space cannot lie outside the circle of radius $t$ centered around $s$ in the scatter plot. 
	}
	\label{fig_simple_scatter}
\end{figure}
It is important to be aware, in this and the following figures, of the importance of the four-point property. The same diagram can of course be plotted for a simple metric space, but in this  case no  spatial relationship is implied between any two points plotted: no matter how close two points are in the plot, there is no implication for the distance between them in the original space. However if  the  diagram is plotted for a metric with the four-point property, then the distance between any two points on the plane is a lower bound on their distance in the original space; two points that are further than $t$ on the plot cannot be within $t$ of each other in the original space. This observation leads to an arbitrarily large number of ways of partitioning the space and allowing these partitions to excluded based on a query position, and   has many potential uses in metric indexing.
\subsection{Indexes Based on Tetrahedral/Planar Projection} \label{sec:Index}
\label{section_core_ideas}
During construction of an index, the constructed 2D space can be arbitrarily partitioned according to any rule based on the geometry of this plane, calculated with respect to the  distances $d(s_i,p_1), d( s_i,p_2)$ and $d(p_1,p_2)$. At query time, if the query falls in any region of the plane that is further than the query threshold $t$ from any such partition, points within that partition cannot contain any solution to the query. 
Since, as will be shown, different spaces give quite different distributions of points within the plane, build-time partitions can be chosen according to this distribution, rather than as a fixed attribute of an index mechanism.

There is much potential for investigating partitions of this plane, and our work is ongoing. The simplest such mechanism to consider is the application of this concept  to normal hyperplane partitioning. Suppose that a data set $S$ is simply divided according to which of the points $p_1$ and $p_2$ is the closer, which corresponds in the scatter diagram to a split over the Y axis. Then at query time, if the corresponding plot position for the query is further than $t$ from the Y axis, no solutions can exist in the subset closer to the opposing reference point. Figure \ref{fig_8_dim_hilbert} shows the same points, but now highlighted according to this distinction. Those drawn in solid, either side of the Y-axis, are guaranteed to be on the same side of the corresponding hyperplane partition in the original space; therefore, if they were query points, the opposing semi-space would not require to be searched. {We refer to these points as ``exclusive queries''.}
\begin{figure*}[tb]
	\centering
	{\includegraphics[width=0.45\textwidth]{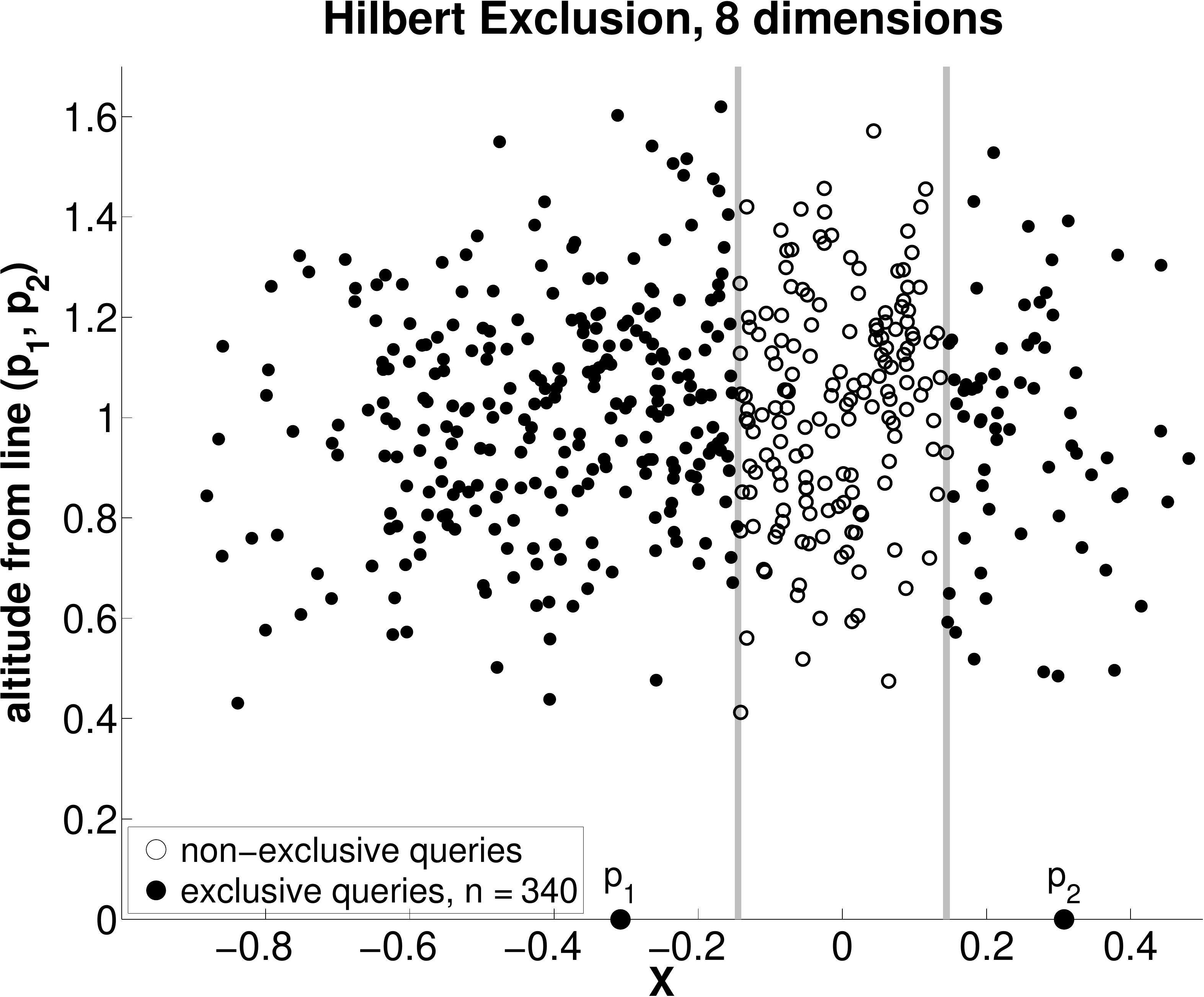}}
	\qquad
	{\includegraphics[width=0.45\textwidth]{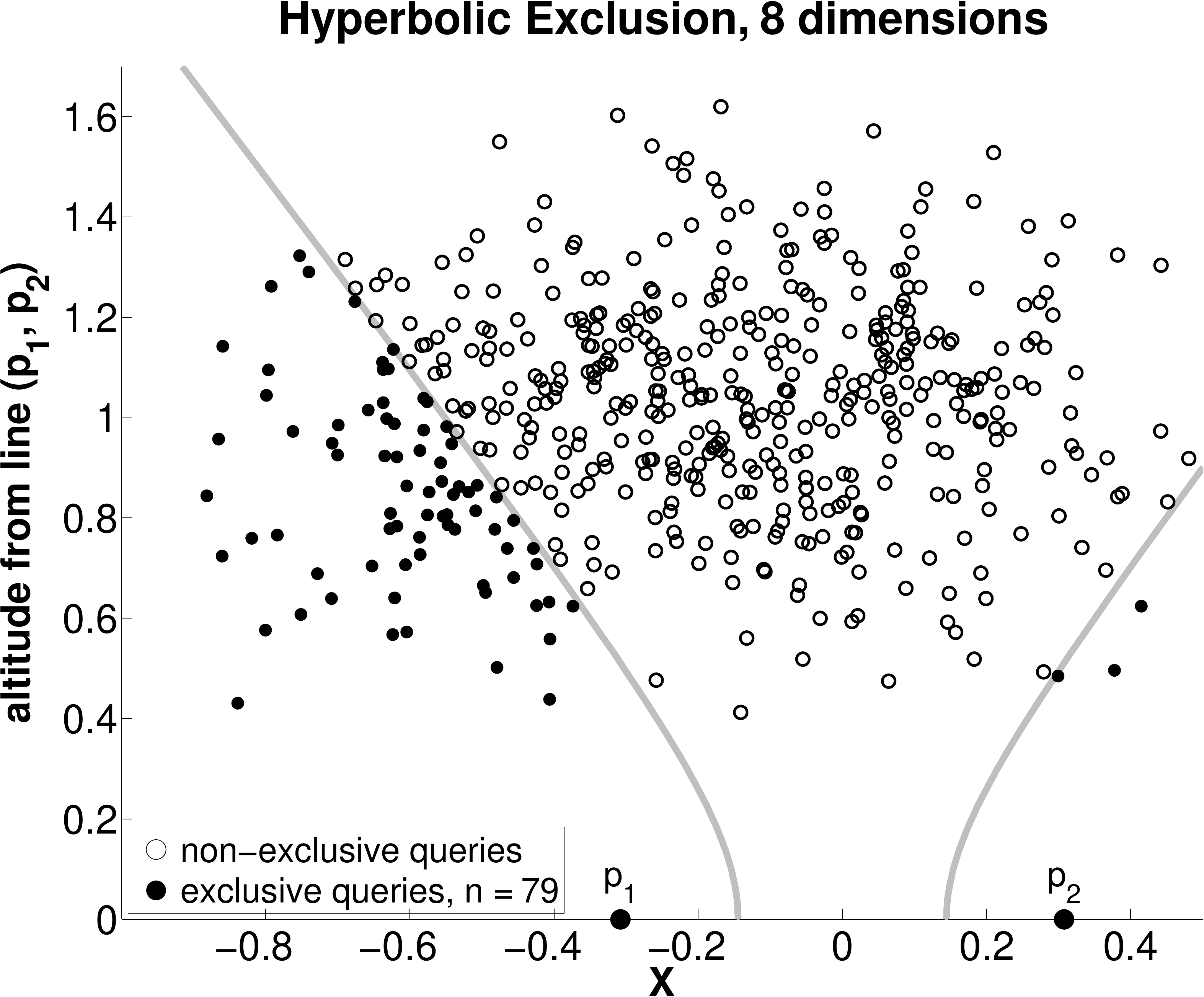}}
	\caption{
		{Scatter diagram for 8-dimensional Euclidean Space.  The data  is divided into two subsets according to which side of Y-axis they lie. The solidly-coloured points are points that, were they queries, would allow the semispace on the opposing side to be excluded from the search since that semispace cannot contain a solution. We refer to these points as ``exclusive queries''. The other points are referred to as ``non-exclusive queries'' since they do not allow the opposing semispace to be excluded from the search.}
		The left-hand side illustrates use of  tetrahedral/planar projection, the right hand side illustrates use of the normal hyperbolic condition.}
	\label{fig_8_dim_hilbert}
\end{figure*}
If the same diagram is drawn for a simple metric space, a query point can be used to exclude the opposing semi-space only according to a  condition algebraically derived from triangle inequality:
$|d(q,p_1) - d(q,p_2)| > 2t $, 
which describes a hyperbola with foci at the reference points and semi-major axis of the search threshold. For the same data and search threshold, the difference in exclusion capability is shown in Figure \ref{fig_8_dim_hilbert}; of the 500 randomly selected queries, only 160  fail to exclude the opposing semi-space, whereas with normal hyperbolic exclusion, the number is 421. The query threshold illustrated, 0.145, is chosen to retrieve around one millionth of the space and is not therefore artificially large.
%

As stated, this particular situation has been extensively investigated and is fully reported in \cite{hilbert_embedding}. Here we will concentrate further on  other properties of the planar projection, of which the derivation of Hilbert exclusion turns out to be a special case.

\subsection{Partitions of the 2D Plane} \label{sec:partitions}
For the purposes of this analysis only, for reasons of simplicity, we seek to divide a data set into precisely two partitions. This is without reference to details of any indexing structure which may use the concepts, although in all cases by implication  there exists a  simple  binary partition tree structure corresponding to the partitioning. In all cases the partition is defined in terms of the 2D plane onto which all points are projected as described above. 

%
\subsection{Reference Point Separation} 
An important observation is that the shape of the 2D ``point cloud", upon which effective exclusion depends, is not greatly affected by the choice of reference points. In comparison with normal Hyperbolic exclusion this is a huge advantage. The hyperbola which bounds the effective queries, i.e. those which can be used to exclude the opposing semispace, is defined only by the (fixed) query radius, and the distance between the reference points, where the larger the separation of the reference points, the better the exclusion. In the extreme case where the separation is no larger than  twice the query radius, which can readily occur in high-dimensional space, it is impossible for any exclusions to be made. This effect can be ameliorated by choosing widely separated reference points, but in an unevenly distributed set this in itself can be dangerous: if one point chosen is an outlier, then the point cloud will lie close to the other point, and  again no exclusions will be made. Finding two reference points which are well separated, and where the rest of the points is evenly distributed between them, is of course an intractable task in general.

Figures \ref{fig_8_dim_far_pivots} and \ref{fig_8_dim_close_pivots} show this effect. In these diagrams, the reference points have been selected as the furthest, and nearest, respectively out of 1,000 sample pairs of points drawn from the space. It can be seen that, when exclusion is based on tetrahedral properties allowed from the four-point property, the exclusive power remains fairly constant, as the size and shape of the point cloud is not greatly affected. However, when the hyperbolic condition is used, the exclusive power is hugely affected; in this case the query threshold is only slightly less than half the separation of the reference points, and the resulting hyperbola diverges so rapidly from the separating hyperplane that no exclusions are made from the sample queries.
\begin{figure*}[tbp]
	\centering
	\includegraphics[width=0.45\textwidth]{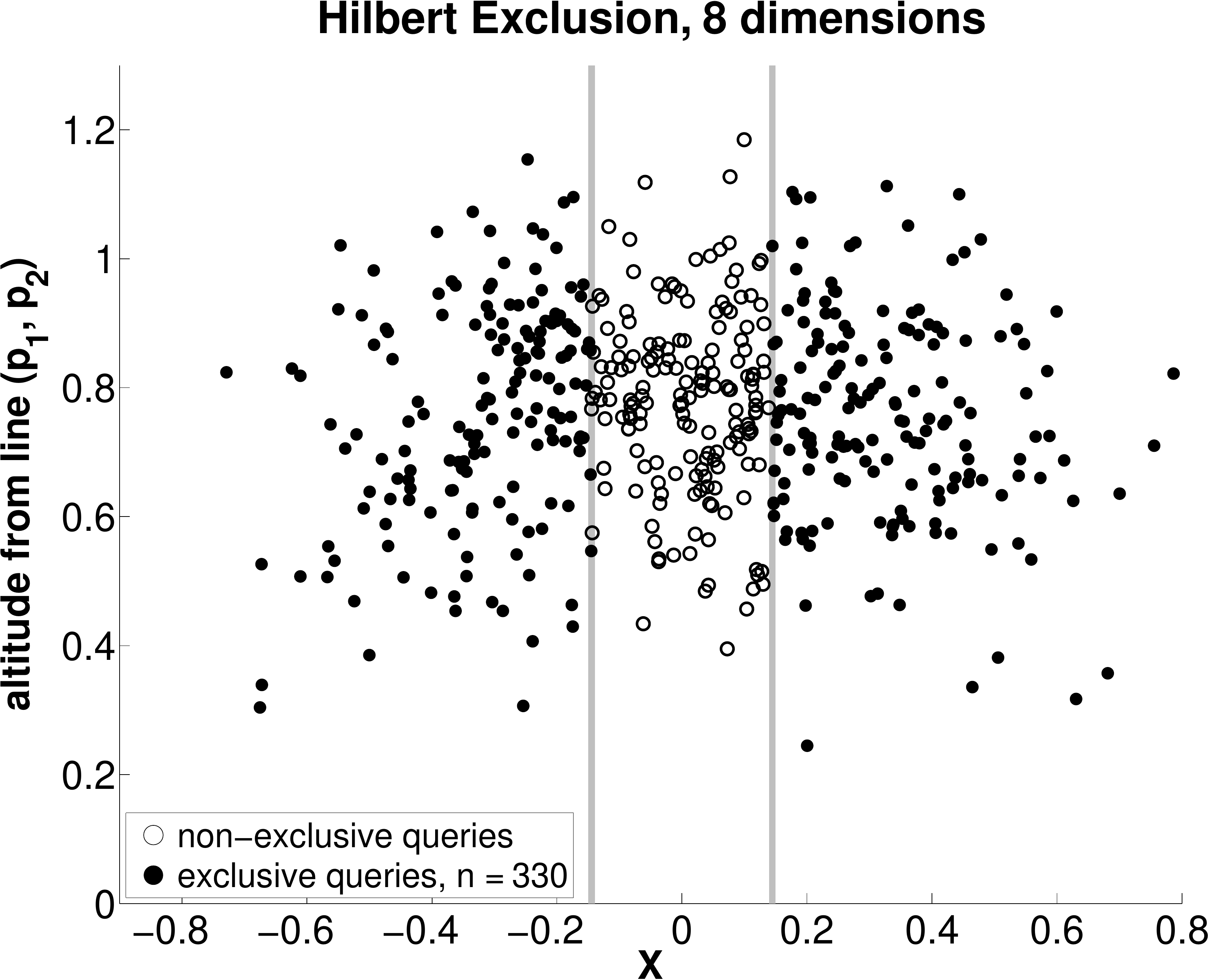}
	\qquad
	\includegraphics[width=0.45\textwidth]{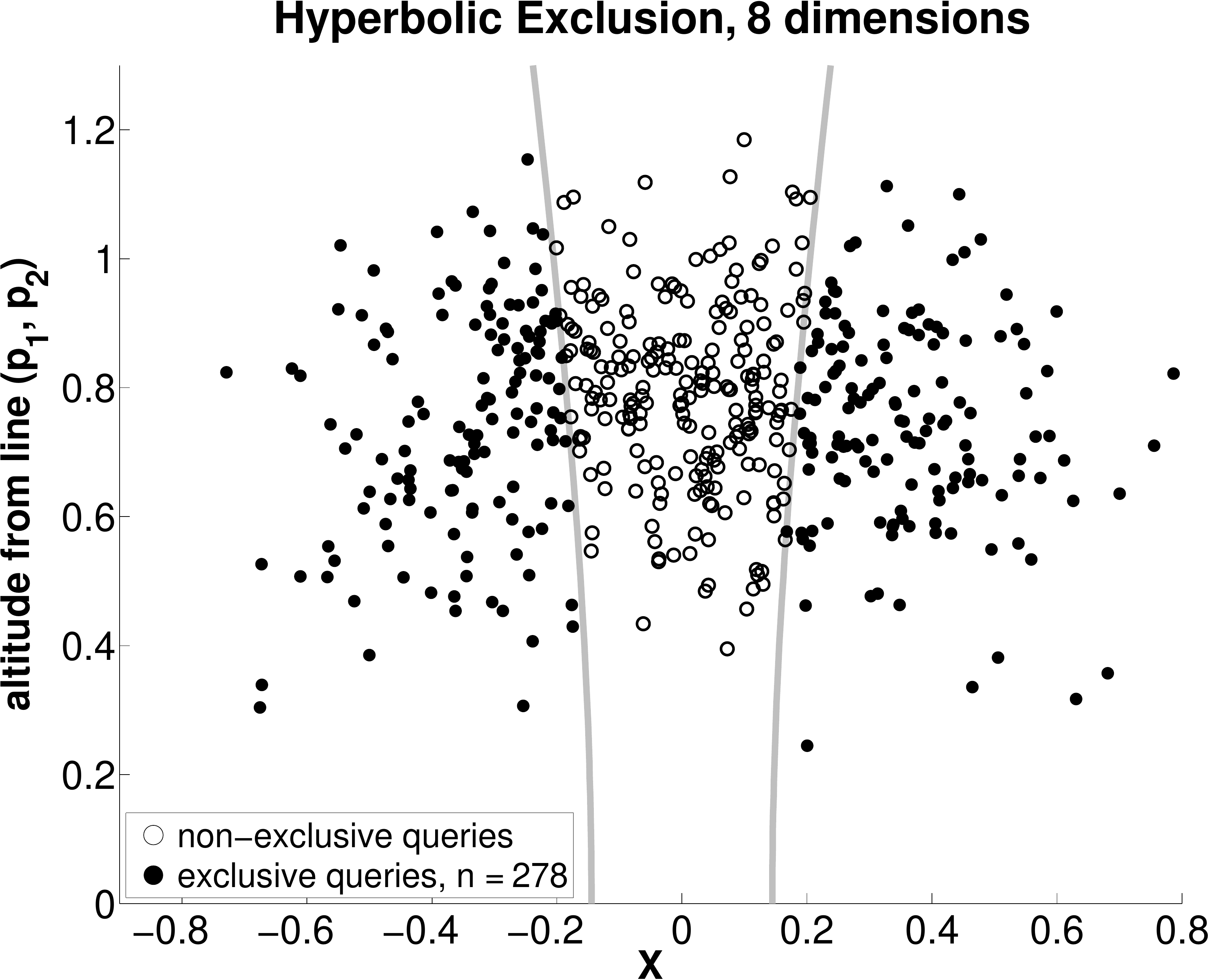}
	\caption{Scatter diagram for 8-dimensional Euclidean Space with widely separated reference points. (The distance between reference points is such that the reference points themselves do not appear on the plot){.}}
	\label{fig_8_dim_far_pivots}
\end{figure*}

\begin{figure*}[tbp]
	\centering
	\includegraphics[width=0.45\textwidth]{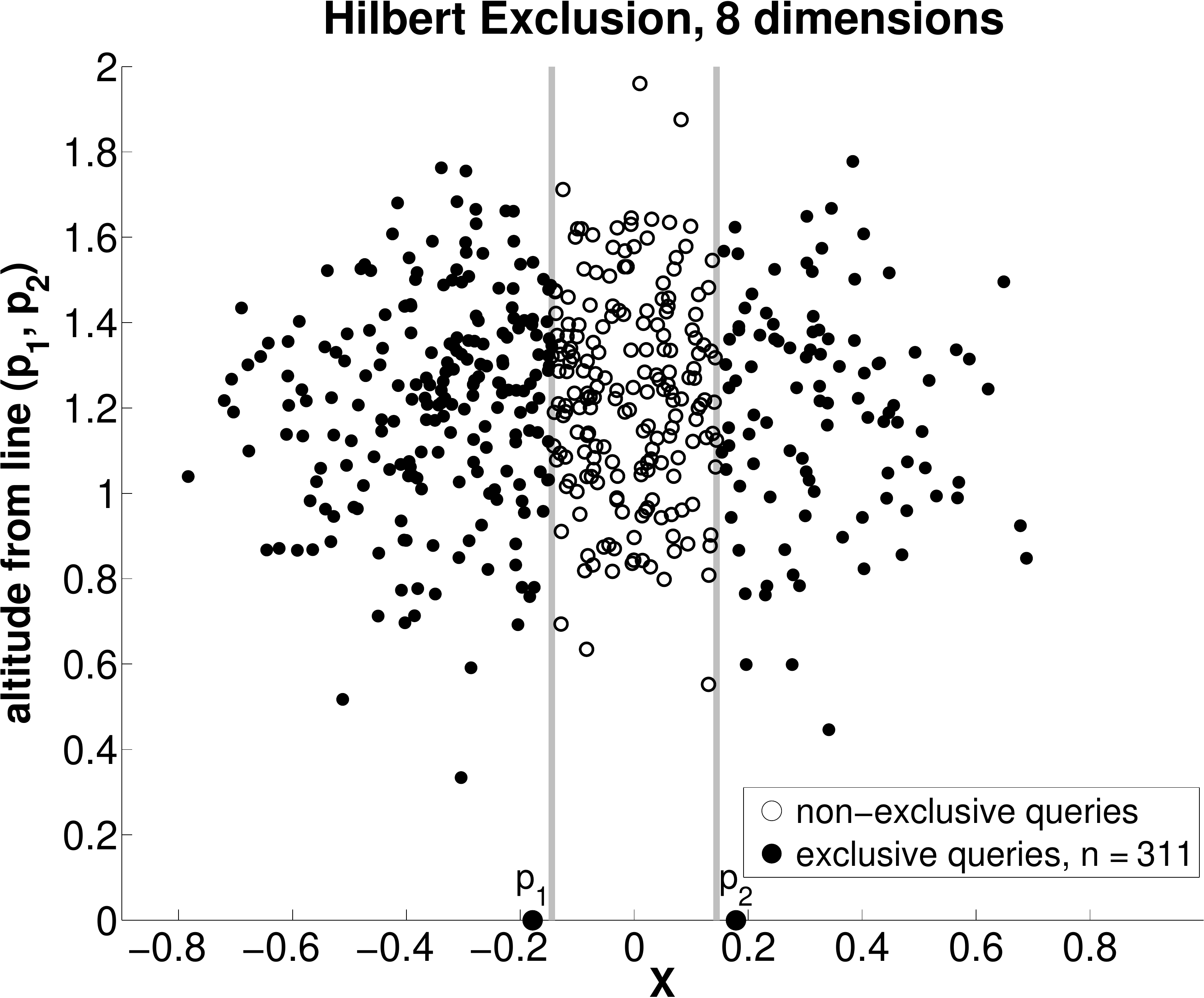}
	\qquad
	\includegraphics[width=0.45\textwidth]{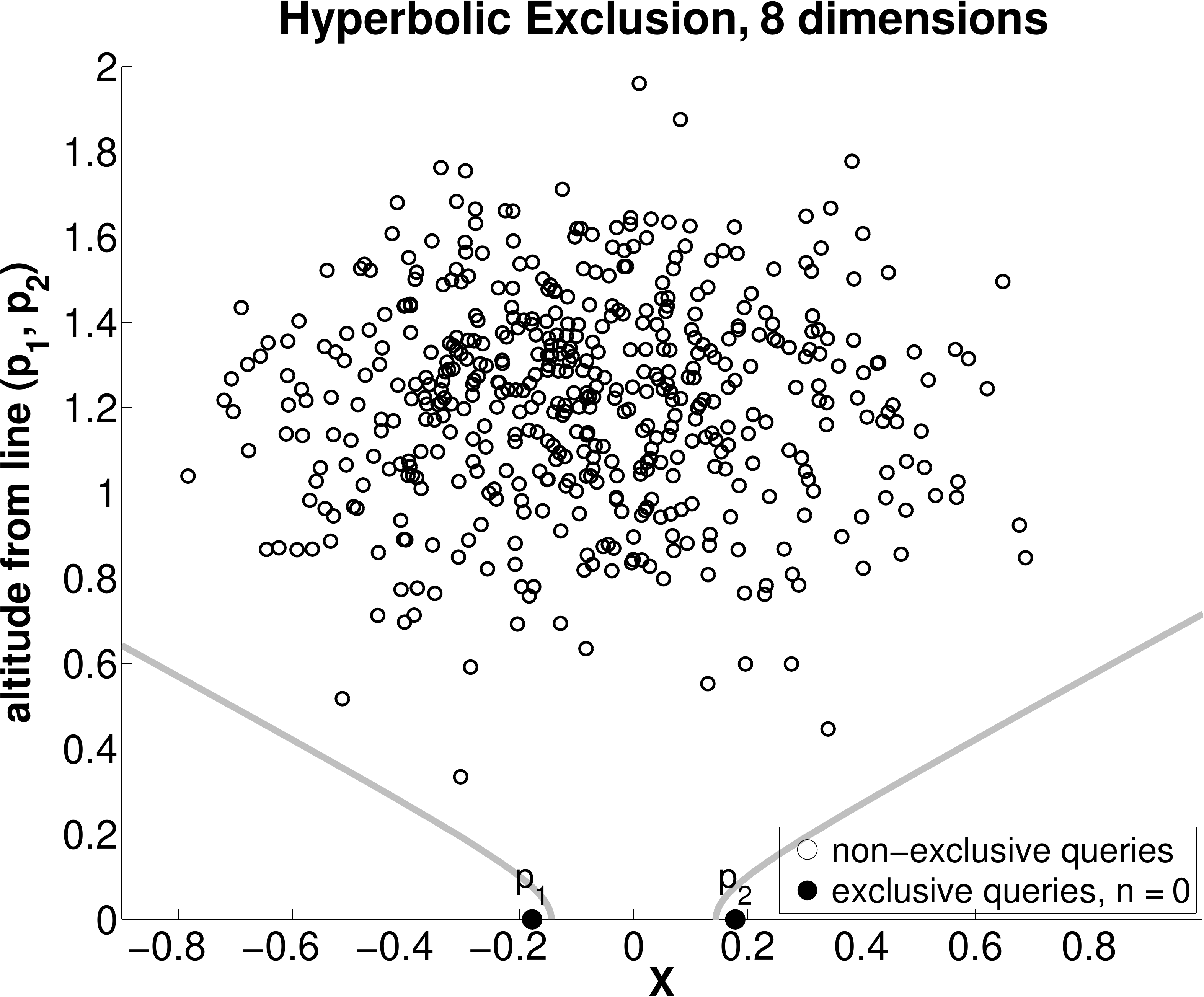}
	\caption{Scatter diagram for 8-dimensional Euclidean Space with close  reference points.
		Note from  comparison of the left-hand graphs of this figure with Figure \ref{fig_8_dim_far_pivots} that the separation of the reference points has no apparent effect on the power of the four-point exclusion, whereas normal metric exclusion becomes completely useless.}
	\label{fig_8_dim_close_pivots}
\end{figure*}
From Figure \ref{fig_8_dim_far_pivots} it should also be noted that, no matter how far the reference points are separated, the four-point property always gives {a higher probability of} 
exclusions; in this case, although the separating lines do not appear visually to be  very different, the implied probability of exclusion in for the four-point property is 0.66, against {0.56}.

To allow most partition structures to perform well, a very large part of the build cost is typically spent in the selection of good reference points and this cost is largely avoidable with any such four-point strategy, as demonstrated experimentally in Sections \ref{sec:exp} and \ref{sec_linear_regression}.
%

\subsection{Arbitrary Partitions}
\label{sec_arbitrary_partitions}
Again we stress the fact that, given the strong lower bound condition on the projected 2D plane, we can choose arbitrary geometric partitions of this plane to structure the data. 
For randomly generated, evenly distributed points there seems to be little to choose. However it is often the case that ``real world" data sets do not show the same properties as generated sets; in particular, they tend to be much less evenly distributed, with significant numbers of clusters and outliers. These factors can significantly affect the performance of indexing mechanisms. 

Figures \ref{fig_colors_alternatives_1}, \ref{fig_colors_alternatives_PCA},  \ref{fig:linear regression}  and \ref{fig_colors_alternatives_2} show a sample taken from the SISAP \emph{colors} data set with Euclidean distance applied, showing eight different partitions. Eight different partitions of the plane have been arbitrarily selected and applied. The query threshold illustrated is 0.052 corresponding to a query returning $0.001\%$ of the data.

\begin{figure*}[tbp]
	\centering
	{\includegraphics[trim=30mm 90mm 35mm 90mm,width=0.46\textwidth]{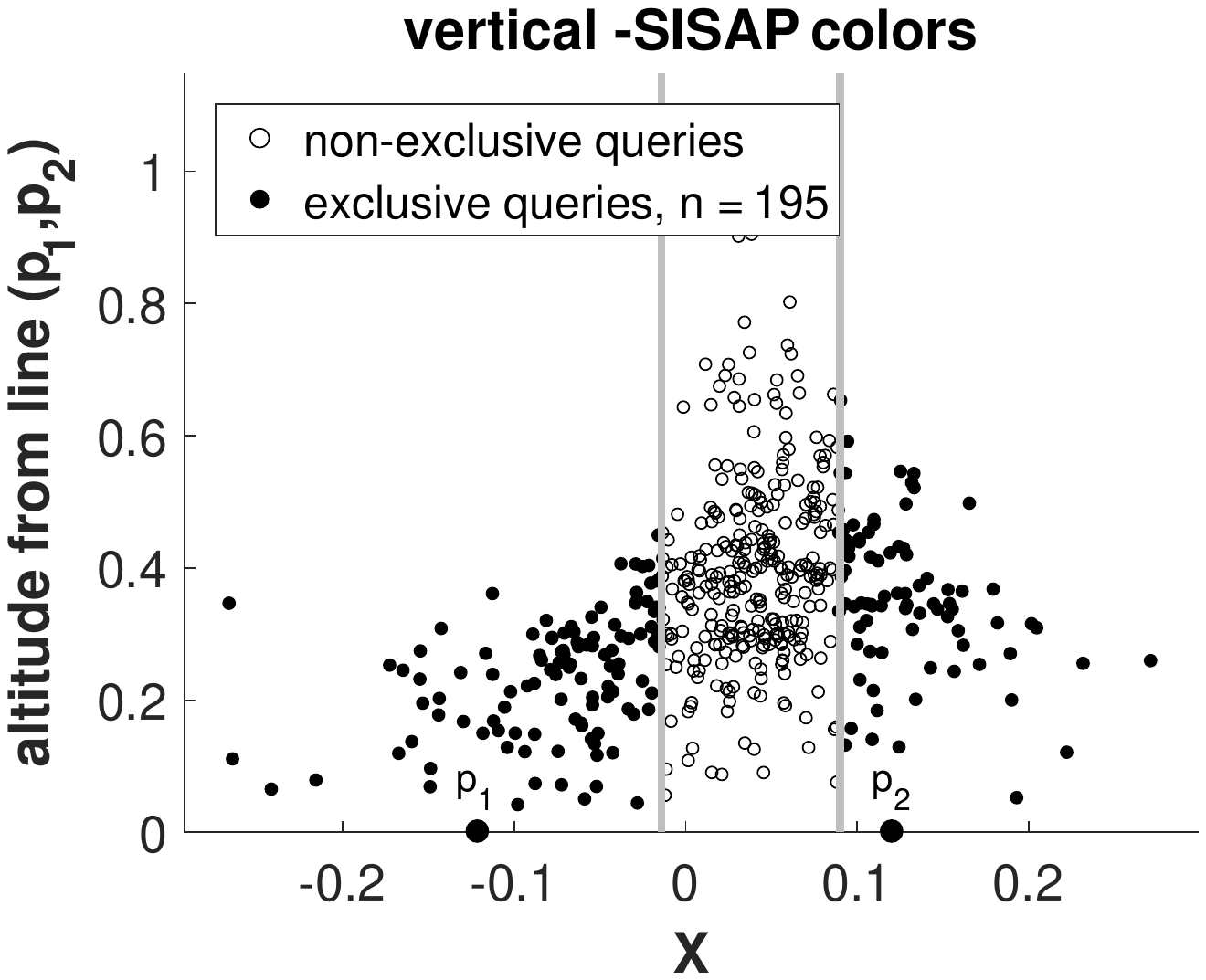}}
	\qquad
	\includegraphics[trim=30mm 90mm 35mm 90mm,width=0.46\textwidth]{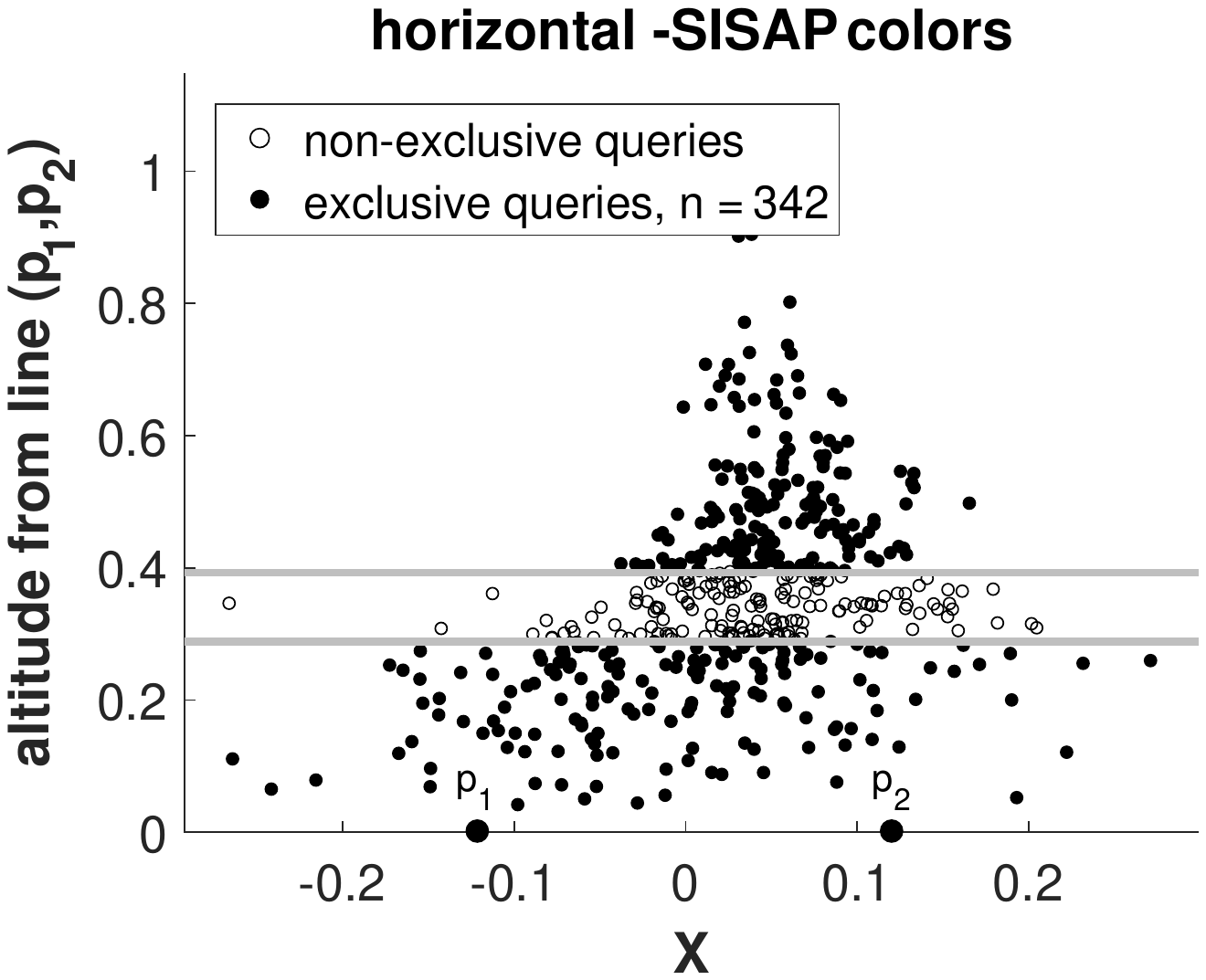}
	\caption{Scatter diagrams dividing the plane equally in X and Y dimension, either can be used for partitioning a hyperplane tree structure; in this case, the horizontal partition would be more effective.}
	\label{fig_colors_alternatives_1}
\end{figure*}

\begin{figure*}[tbp]
	\centering
	{\includegraphics[trim=30mm 90mm 35mm 90mm,width=0.46\textwidth]{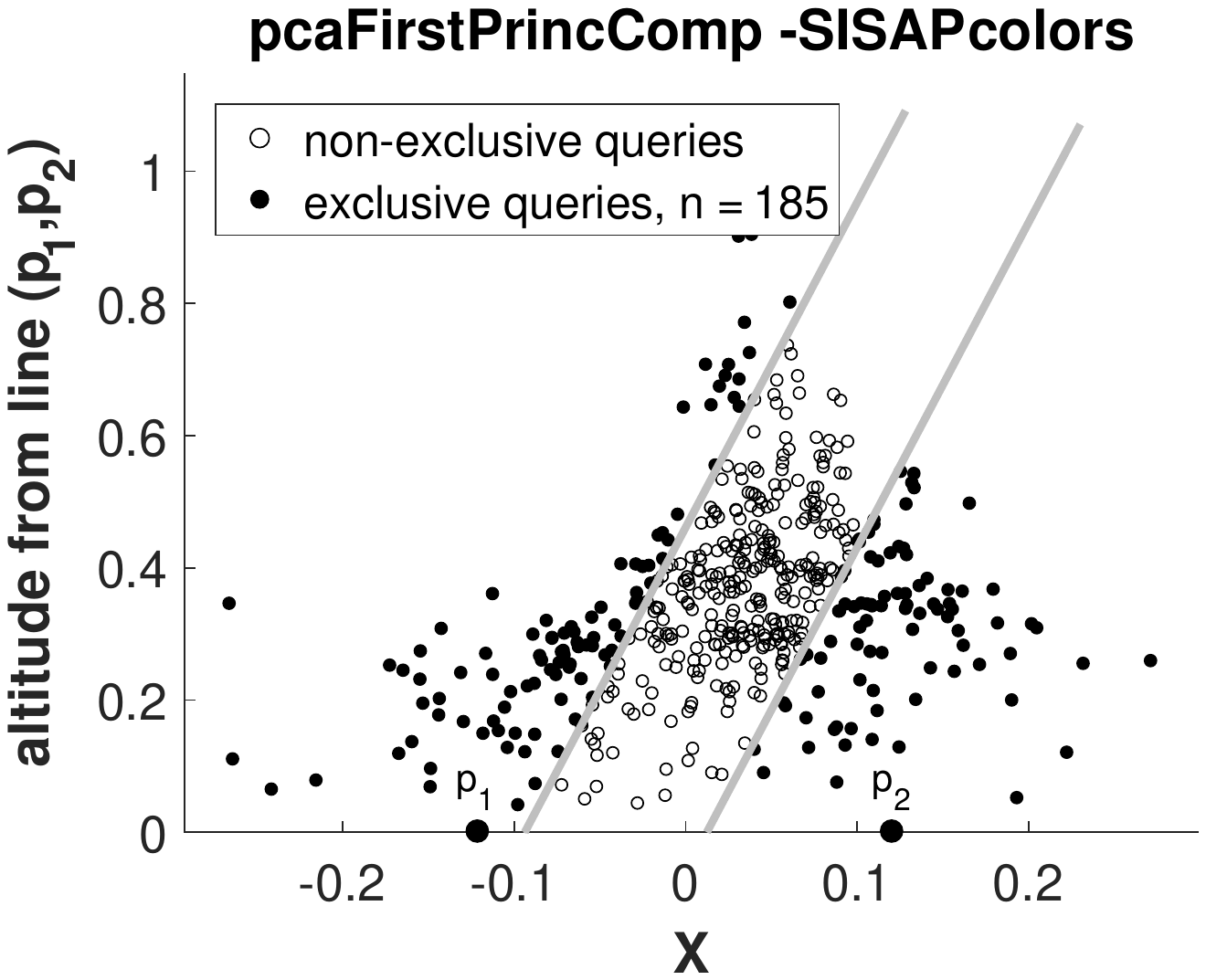}}
	\qquad
	\includegraphics[trim=30mm 90mm 35mm 90mm,width=0.46\textwidth]{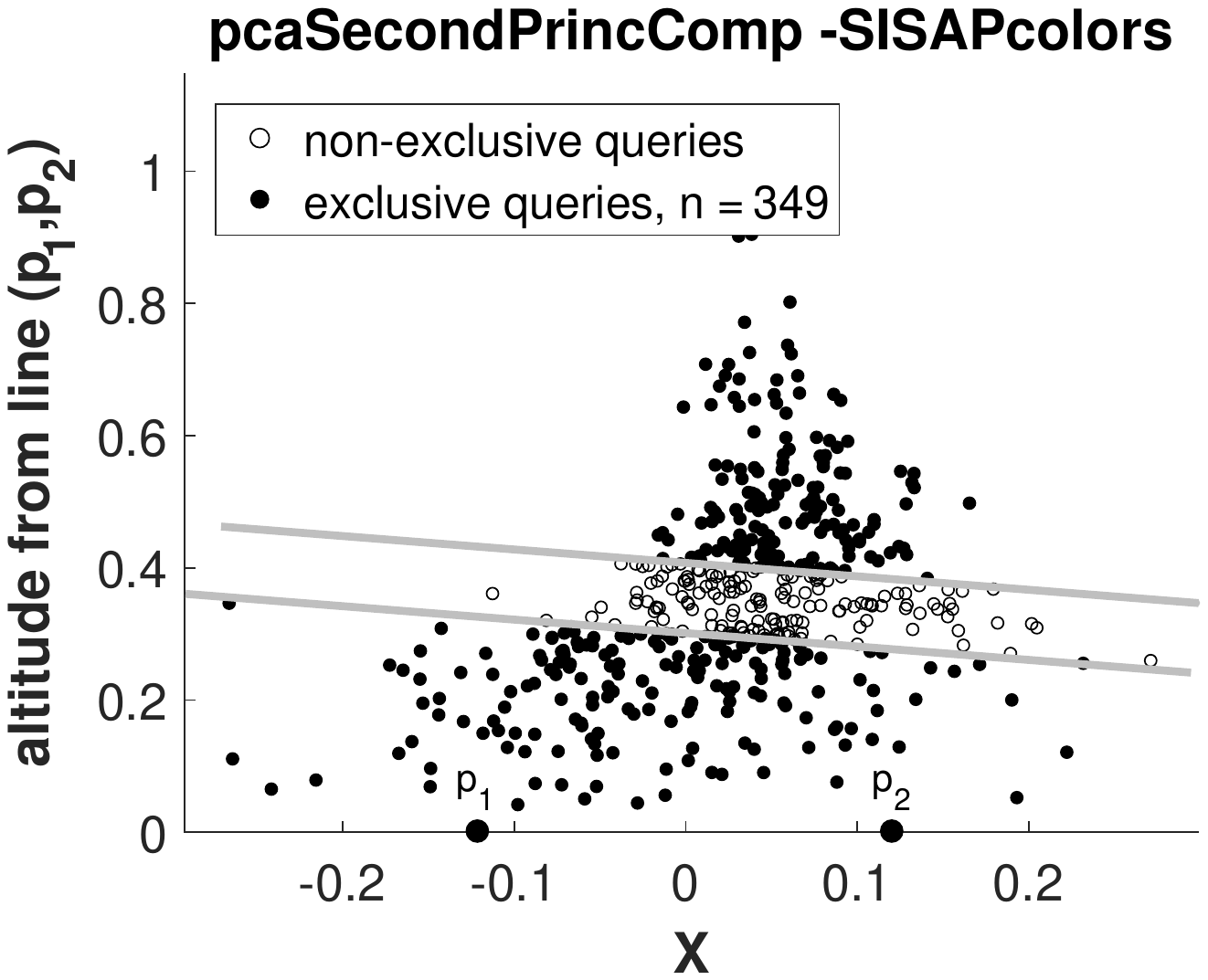}
	\caption{Hyperplane partitioning based on a hyperplane parallel to the first (left) and the second (right) principal components.}
	\label{fig_colors_alternatives_PCA}
\end{figure*}
\begin{figure*}[tbp]
	\centering
	{\includegraphics[trim=30mm 90mm 35mm 90mm,width=0.46\textwidth]{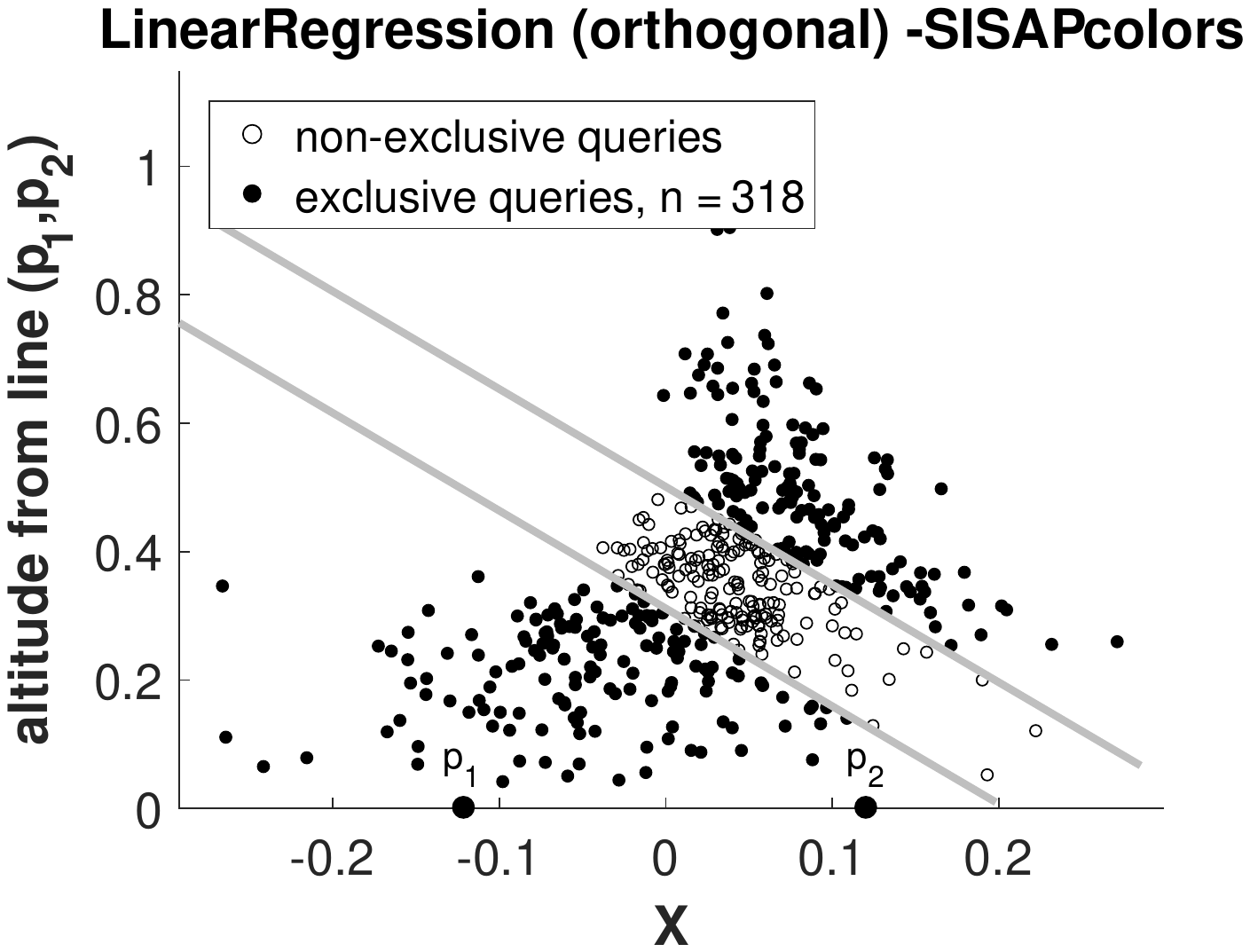}}
	\qquad
	{\includegraphics[trim=30mm 90mm 35mm 90mm,width=0.46\textwidth]{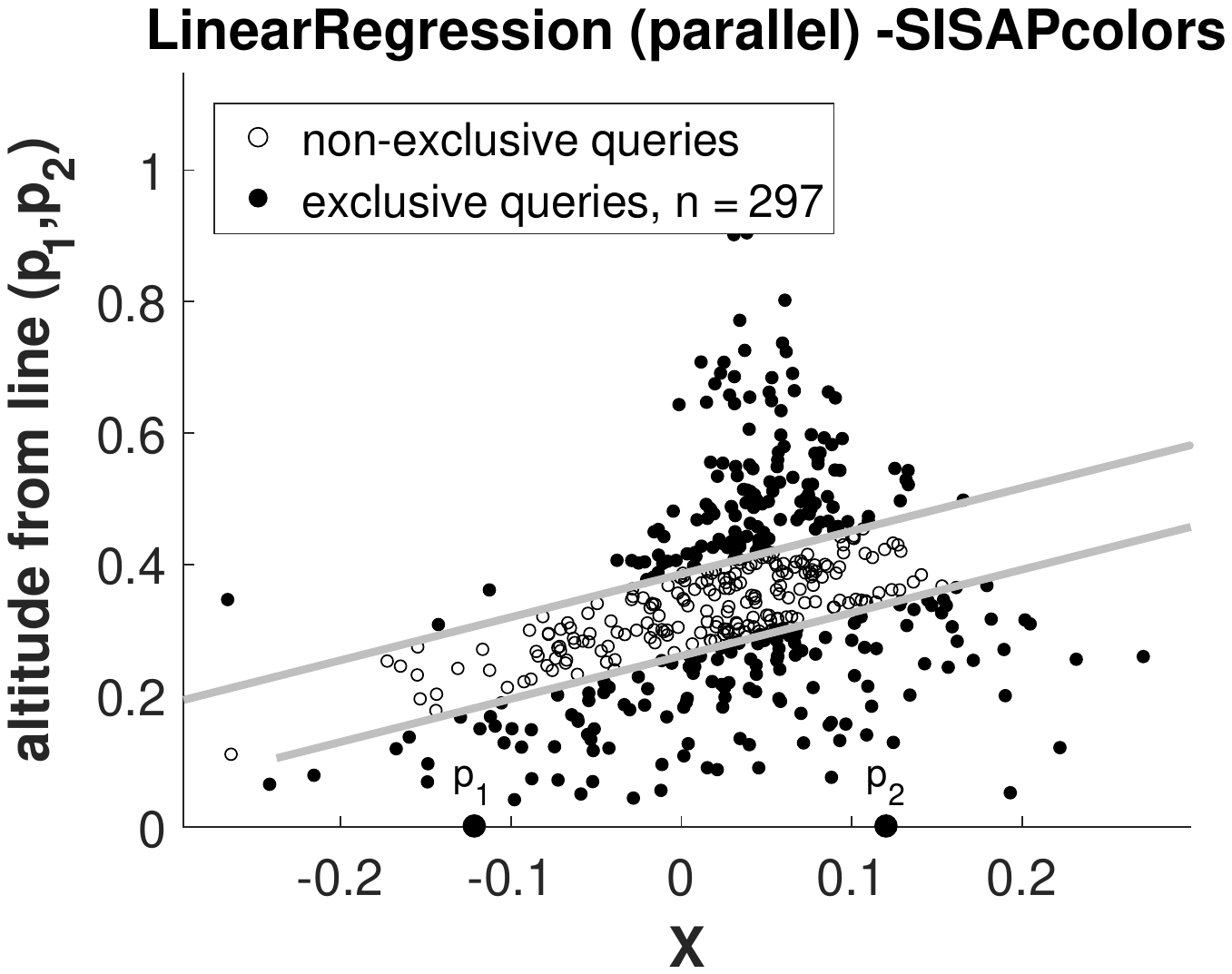}}
	\caption{Hyperplane partitioning based on a hyperplane orthogonal (left) or parallel (right) to the best-fit line through data.}
	\label{fig:linear regression}
\end{figure*}
\begin{figure*}[tbp]
	\centering
	{\includegraphics[trim=30mm 90mm 35mm 90mm,width=0.46\textwidth]{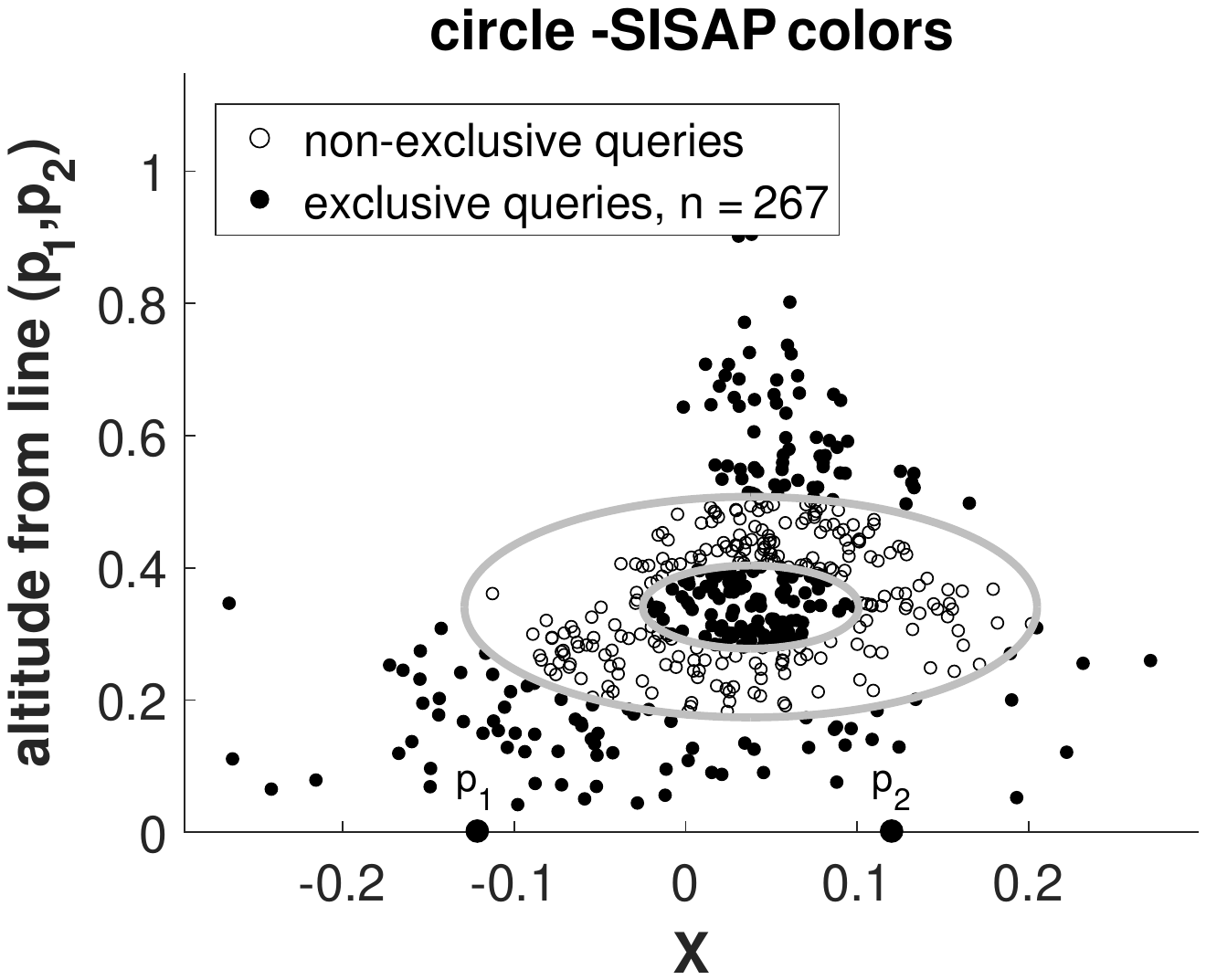}}
	\qquad
	\includegraphics[trim=30mm 90mm 35mm 90mm,width=0.46\textwidth]{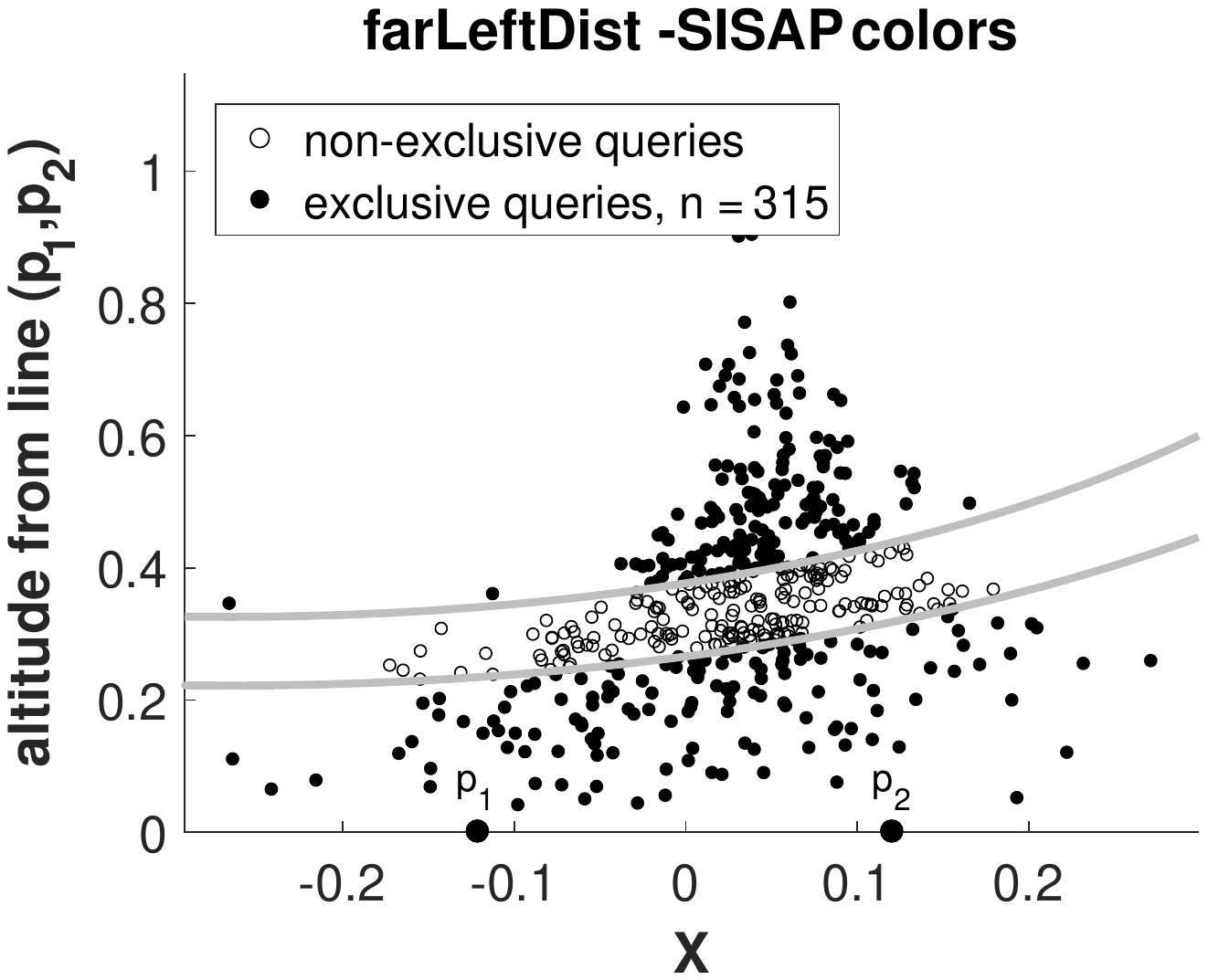}
	\caption{Two more  binary partitions, based now on median distance from arbitrary points in the plane (centre and top-left respectively); we have not yet found a use for these but include the diagrams to make the point that any such partition may be used.}
	\label{fig_colors_alternatives_2}
\end{figure*}
In all cases, it can be noted that the partitions are even, leading to balanced indexing structures. It is very likely that skewed partitions may perform better, an aspect we have not yet investigated. However one important balanced partition is illustrated on the left hand side of Figure \ref{fig_colors_alternatives_1}, implying that a balanced hyperplane tree can be efficiently constructed.

It can be seen that, in this case, partitioning the plane according to the height of individual points above the X-axis is the more effective strategy. The disadvantage with this is that a little more calculation is required to plot the height of the point, rather than its offset from the Y-axis; however this is a very minor effect when significantly more distance calculations can be avoided.

Figures \ref{fig_colors_alternatives_PCA} and \ref{fig:linear regression} illustrate more techical analyses of the point cloud, using Principle Component Analysis (PCA) and Linear Regression (LR) respectively. Either technique can be used along one of two axes in a two-dimensional space as illustrated. In Section \ref{sec_linear_regression} we explain in the orthogonal linear regression technique in detail, and give   experimental results showing its value as the best way to construct a balanced search tree over this data.

Finally we give the illustrations in Figure \ref{fig_colors_alternatives_2} to make the points that any partition of the plane can be used for this purpose. We have not yet found a compelling use for either partition, however this would depend on the nature of an individual non-uniform data set.

\subsection{Balance}
As already noted, any of the partitions shown above can be simply used to bisect the data and thus produced a balanced indexing structure. These examples are all defined using a single real value with respect to the planar geometry. This can be calculated for each object within the subset to be divided, and the median can be found very efficiently using the QuickMedianSort algorithm; for a random distribution of points, the practical cost of balancing a binary tree at construction time  appears similar to performing QuickSort once on all the data. While balanced structures are often slower than unbalanced ones for relatively small data sets, they become rapidly more desirable as the size of the data increases, and again more so if it is too large to fit in main memory and requires to be stored in backing store pages. The ability to balance the data without reducing the effectiveness of the exclusion mechanism therefore seems important. 

One further area of investigation, not yet performed, would be the effect of controlling the balance, which once again is arbitrarily possible simply by selecting different offset values. In general this will increase the probability of exclusion at cost of excluding smaller subsets of the data, and the effectiveness will depend on the individual distributions of the different strategies.

\section {The \ourTree}
\label{sec_linear_regression}

In this Section we revisit a key observation of Section  \ref{sec:Proj}, and in particular Section \ref{sec_arbitrary_partitions}, where we pointed out that any partition of the two-dimensional projected plane may be used to form an indexing mechanism. Up to this point we have restricted the use to simple Hilbert partitioning, where the data is divided only according to the nearest reference point. Here, we demonstrate a more flexible approach.

Figure \ref{fig:linear regression} shows
a scatter plot resulting from an arbitrary choice of reference points for the SISAP \emph{colors} data set. Although the pattern is not atypical, observation shows that the individual distribution shape is significantly affected by the choice of reference points and, more subtly, by the subset of data points that is to be stored at a given tree node; although a high-dimensional space implies that these would not necessarily have a strong regional identity, this factor does visibly affect the relative mean distances to the reference points.

The partitions shown within the figure are based on the best-fit straight line which can be plotted through the points in two dimensions. This is parallel to the lines drawn in the right-hand figure. As this is calculated using the least-mean-squares algorithm, it is reasonable to assume that the perpendicular partition, shown in the left-hand diagram, will in general improve the spread of the data points and thus form a better partition for indexing%
\footnote{As shown above, PCA seems to give a better spread than LR; for the moment we have selected linear regression for the experiment due primarily to its simplicity of implementation, we continue to investigate alternative strategies.}.

To test this strategy, we define the {\ourTree}\, ({\oT}), which is a binary tree built recursively over a dataset $S$ as follows. We select two reference points $p_1$, $p_2$ at each node. Each child node of the tree shares one reference points with its parents, as done in the Monotonous Bisector Tree~\cite{Noltemeier1992}. We used the tetrahedral projection based on $p_1$ and $p_2$ to embed the data points onto a 2D plane, and we compute the best-fit line $l$ through the projected points (or a subset of them) using a least squares minimization. Then, we rotate the 2D data points around the X-intercept of the line $l$, so that the new X-axis coincides with the line $l$, and we split the data  at the median X coordinate of the rotated space.

Algorithm \ref{createnode} and \ref{query} give the simplest algorithms for constructing, and querying a balanced version of the  \oT.  

We compute the best fitting line $l$ through the points $\{(x_i,y_i)\}_{i=1}^N$ as the line $y=mx +b$ that best fits the sample in the sense that the sum of the squared errors between the $y_i$ and the line values $mx_i+b$ is minimized. The fitting line is easily computed as  $y-\bar{y}=m(x-\bar{x})$ where $\bar{x}=\sum_{i=1}^N x_i/N$,  $\bar{y}=\sum_{i=1}^N y_i/N$, and 
\begin{equation}
m=\dfrac{\sum_{i=1}^N(x_i-\bar{x})(y_i-\bar{y})}{\sum_{i=1}^N(x_i-\bar{x})^2}.
\end{equation}

Then, we rotate the data points by angle $\theta=\arctan(m)$ around the X-intercept $(h,0)$, where $h=\bar{x}-\bar{y}/m$:
\begin{align}
r_x&= (x-h) \cos(\theta)-y \sin(\theta)\\
r_y&=(x-h) \sin(\theta)+y \cos(\theta).
\end{align}


\begin{algorithm}[tp]
	\small{
		\SetKwInOut{Input}{Input}
		\SetKwInOut{Output}{Output}
		\Input{$A\subset S,\, p_1\in S$}
		\Output{{Node: $N=\left\langle p_1,p_2,\, \delta, \, \theta,\, h,\,N_\text{left}, N_\text{right}\right\rangle$ where $\{p_1,p_2\}\subset U,\, \delta \in  \R{},\, \theta \in [0,2\pi),\,  h\in \R{}, \{N_\text{left}, N_\text{right}\}\subset \text{Node}$}}
		Select $p_2$ from $A$\;
		\If{$|A|>2$}{
			$A\gets A\smallsetminus\{p_1,p_2\}$ \;
			$\widetilde{A} \gets$2Dproject{($A,p_1,p_2$)}\; 
			$(\theta,h)\gets$GetRotationAngle($\widetilde{A}$)\tcp*[l]{Calculate the rotation angle $\theta$, and the X-intercepts  $(h,0)$, that minimize the squared errors of the $y$-coordinates following the rotation transformation} 
			$RotatedPoints\gets \emptyset$\;
			\ForEach{ $\widetilde{s_j}$ in $\widetilde{A}_i$}{ $ r_j\gets  \text{Rotate}(\widetilde{s_j}, \theta, h)$ \tcp*[l]{ $r_j=({r_j}.x,\,{r_j}.y) \in \R{2}$
				}
				$RotatedPoints\gets RotatedPoints\cup \{r_j\}$	
				
			}
			$\delta\gets \text{median}\{r_j.x|\, r_j\in RotatedPoints \}$ \tcp*[l]{Find the median value of the $x$-coordinate of the rotated points}
			$A_\text{left} \gets \{ s_j\in A|\, {r_j}.x,< \delta \}$\;
			$A_\text{right}\gets\{ s_j\in A|\, {r_j}.x,\geq \delta \}$\;
			$N_\text{left} \gets$ CreateNode($A_\text{left}$, $p_1$)\;
			$N_\text{right} \gets$ CreateNode($A_\text{right}$, $p_1$)\;
			$N \gets \left\langle p_1,p_2,\, \delta, \, \theta,\, h,\,N_\text{left},
			N_\text{right}\right\rangle$\;
	}}
	\caption{{CreateNode} ({\oT} balanced)}\label{createnode}
\end{algorithm}

\begin{algorithm}[tp]
	\small{
		\SetKwInOut{Input}{Input}
		\SetKwInOut{Output}{Output}
		\Input{$A\subset S,\, p_1, p_2\in S$}
		\Output{Set $\widetilde{A}\subset \R{2}$}
		$\widetilde{A}\gets \emptyset$\;
		\ForEach{ $s_j$ in $A$}{
			Calculate the 2D embedded point $\widetilde{s_j}$ as the apex of the triangle defined by baseline $(0,-d(p_{1},p_{2})/2) - (0,d(p_{1},p_{2})/2)$, with left side length $d(s_j,p_{1})$ and right side $d(s_j,p_{2})$\;
			$\widetilde{A}\gets \widetilde{A} \cup\{\widetilde{s_j}\}$\;	}}
	\caption{{2Dproject} (2D projection of $A$ based on $p_1,\, p_2$)}\label{2Dproj}
\end{algorithm}

\begin{algorithm}[tp]
	\small{
		\SetKwInOut{Input}{Input}
		\SetKwInOut{Output}{Output}
		\Input{$q\in U,\, t\in \R{},\, N=\left\langle p_1,p_2,\, \delta, \, \theta,\, h,\,N_\text{left}, N_\text{right}\right\rangle\in$ Node}
		\Output{Result set $R=\{s\in S| \, d(s,q)\leq t\}$}
		$R \gets \emptyset$\;
		\If{$d(q,\, p_1)\leq t$}{
			$R \gets R \cup \{p_1\}$\;
		}
		\If{$d(q,\, p_2)\leq t$}{
			$R \gets R \cup \{p_2\}$\;
		}
		$\widetilde{q}\leftarrow$ 2Dproject$(\{q\},p_1,p_2)$\;  
		$r_q= \text{Rotate}(\widetilde{q}$, $\theta$, $h$)\;
		\eIf{${r_q}.x  < \delta - t $}{
			$R \gets R \cup Query(q, N_\text{left})$\;
		} 
		{	\eIf{${r_q}.x  > \delta + t $}{
				$R \gets R \cup Query(q, N_\text{right})$\;}
			{
				$R \gets R \cup Query(q, N_\text{left})$\;
				$R \gets R \cup Query(q, N_\text{right})$\;
			}
	}}     	     
	\caption{{Query}}\label{query}
	
\end{algorithm}

Experimental evaluation of the resulting search index was performed using exactly the same context as that described in Section \ref{sec:exp}, and   all of the code used is available from the same repository. Figures \ref{fig:LRTcolor} and \ref{fig:LRTnasa} give results for the SISAP \emph{colors} and \emph{nasa} data sets respectively. For each data set, six different indexing structures were tested. A balanced monotone hyperplane tree,  an unbalanced monotone hyperplane tree, and the Linear Regression Tree were each tested with two different reference point selection strategies. These are: ``Rand" -- random selection --  and ``Far" --  in the monotone tree, one reference point is handed down from an ancestor, and the second point is simply the one from within the data subset used to construct that node that is the furthest distance from the ancestor node.

The fair comparison is of the two balanced trees, and it can be seen that the Linear Regression Tree always outperforms the simple balanced tree. 

The unbalanced tree however is always the best performer over this data set. Reasons for this are not altogether clear. However we believe this is worth reporting for sake of further investigation: in this domain, successful analysis of these reasons should lead to the ability to mimic them and deterministically produce a tree with still better performance.

In some of the further experiments performed in Section \ref{section_large_experiments}  we find that the Linear Regression Tree performs best out of all the mechanisms tested. This seems to be for large data sets which have significant non-uniformity, searching with smaller thresholds.

\begin{figure*}[t]
	\centering
	
	\includegraphics[trim=30mm 90mm 30mm 90mm,width=0.45\textwidth]{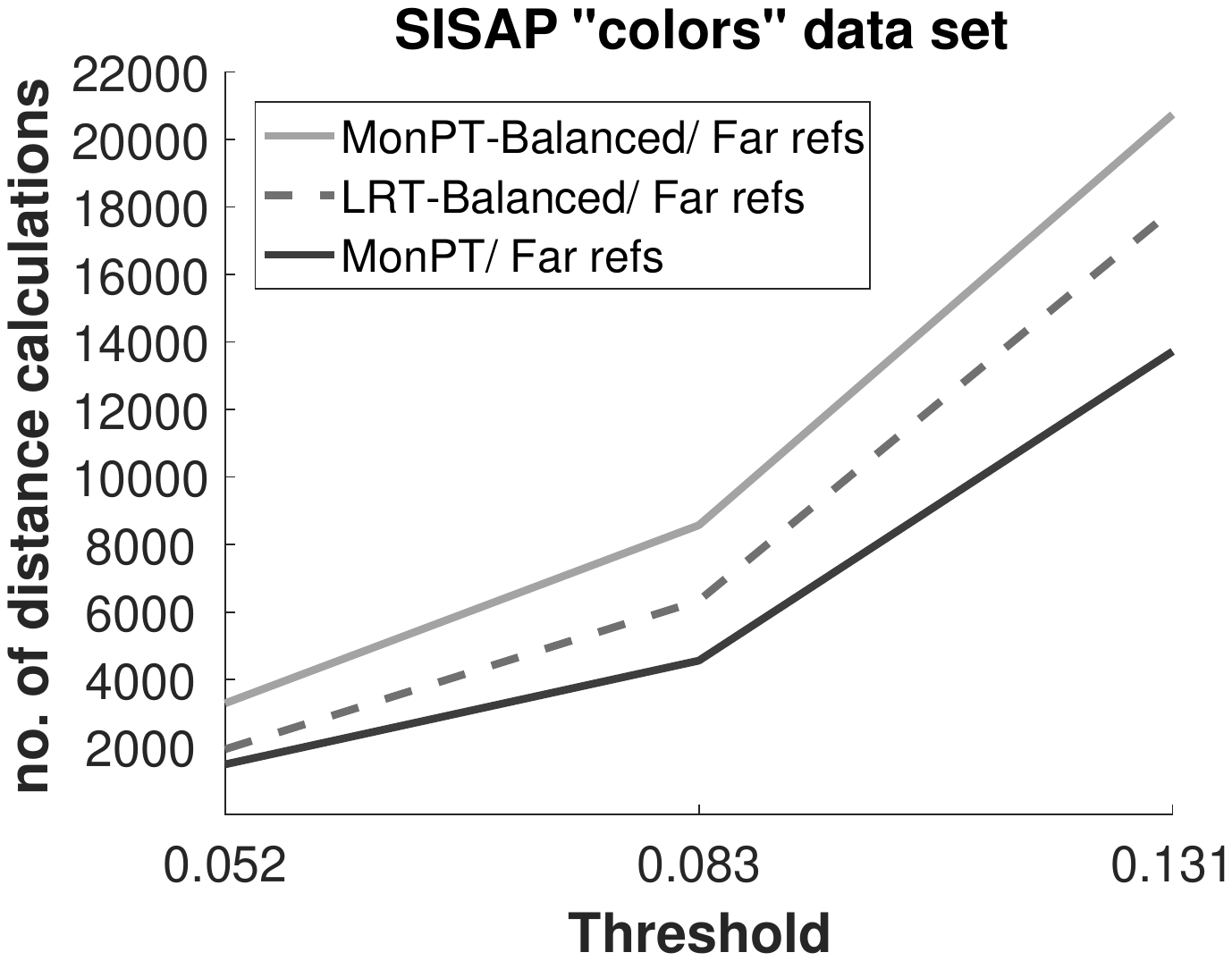}
	\qquad
	\includegraphics[trim=30mm 90mm 30mm 90mm,width=0.45\textwidth]{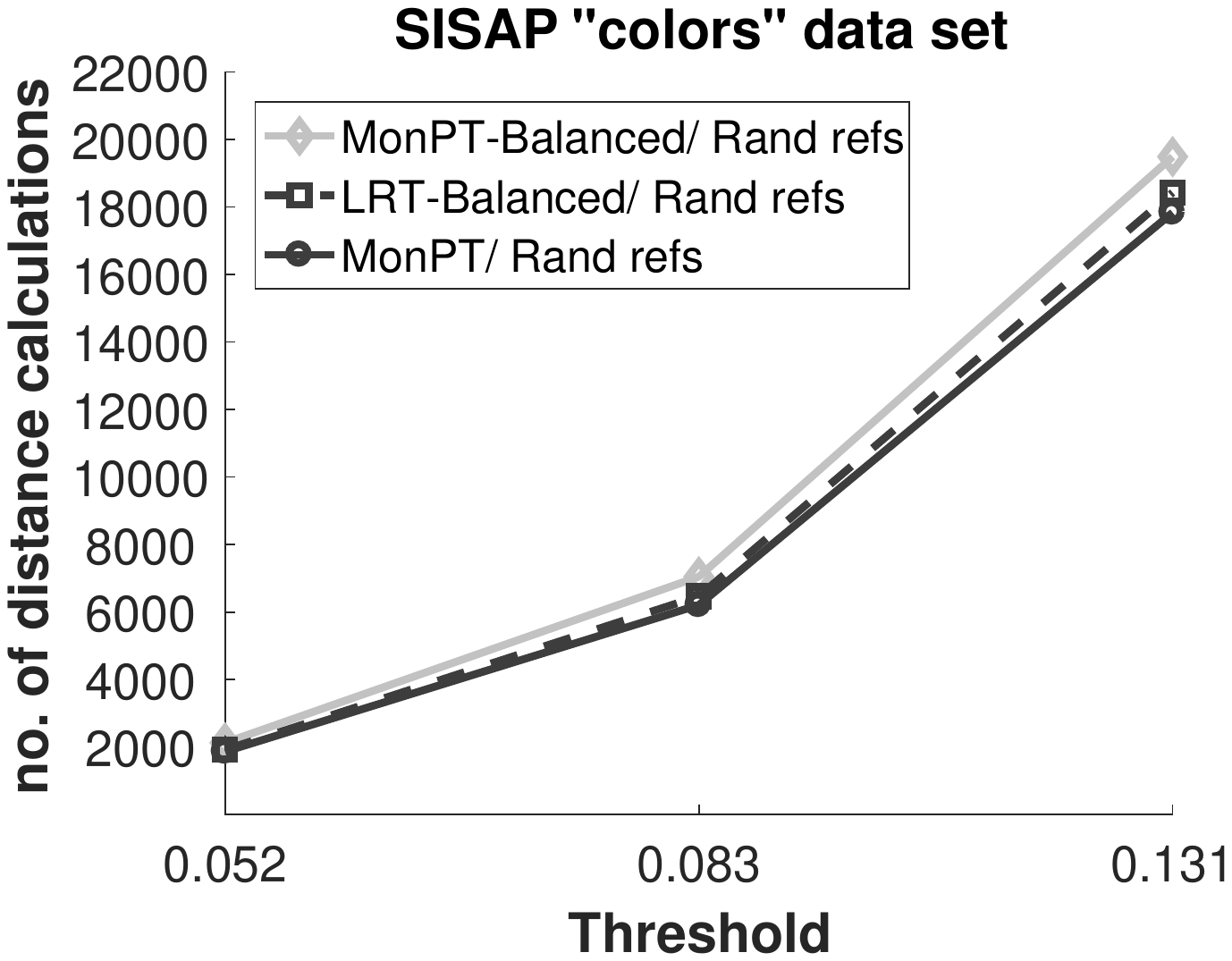} 
	\caption{Colors --Monotone Tree (MonPT), Linear Regression Tree (LRT), and Balanced Monotone Tree with two different reference point selection strategies}
	\label{fig:LRTcolor}
\end{figure*}

\begin{figure*}[t]
	\centering
	
	\includegraphics[trim=30mm 90mm 30mm 90mm,width=0.45\textwidth]{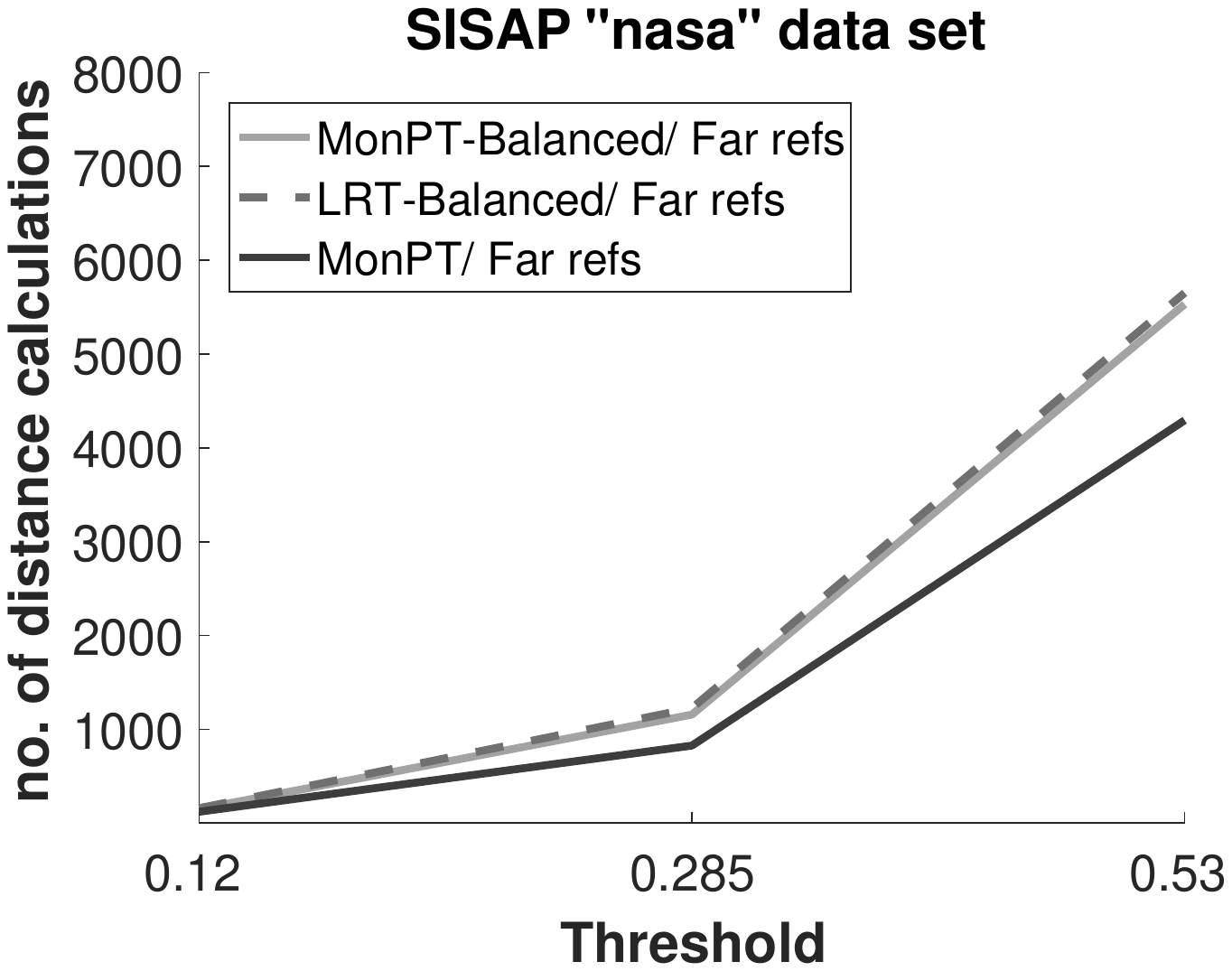}
	\qquad
	\includegraphics[trim=30mm 90mm 30mm 90mm,width=0.45\textwidth]{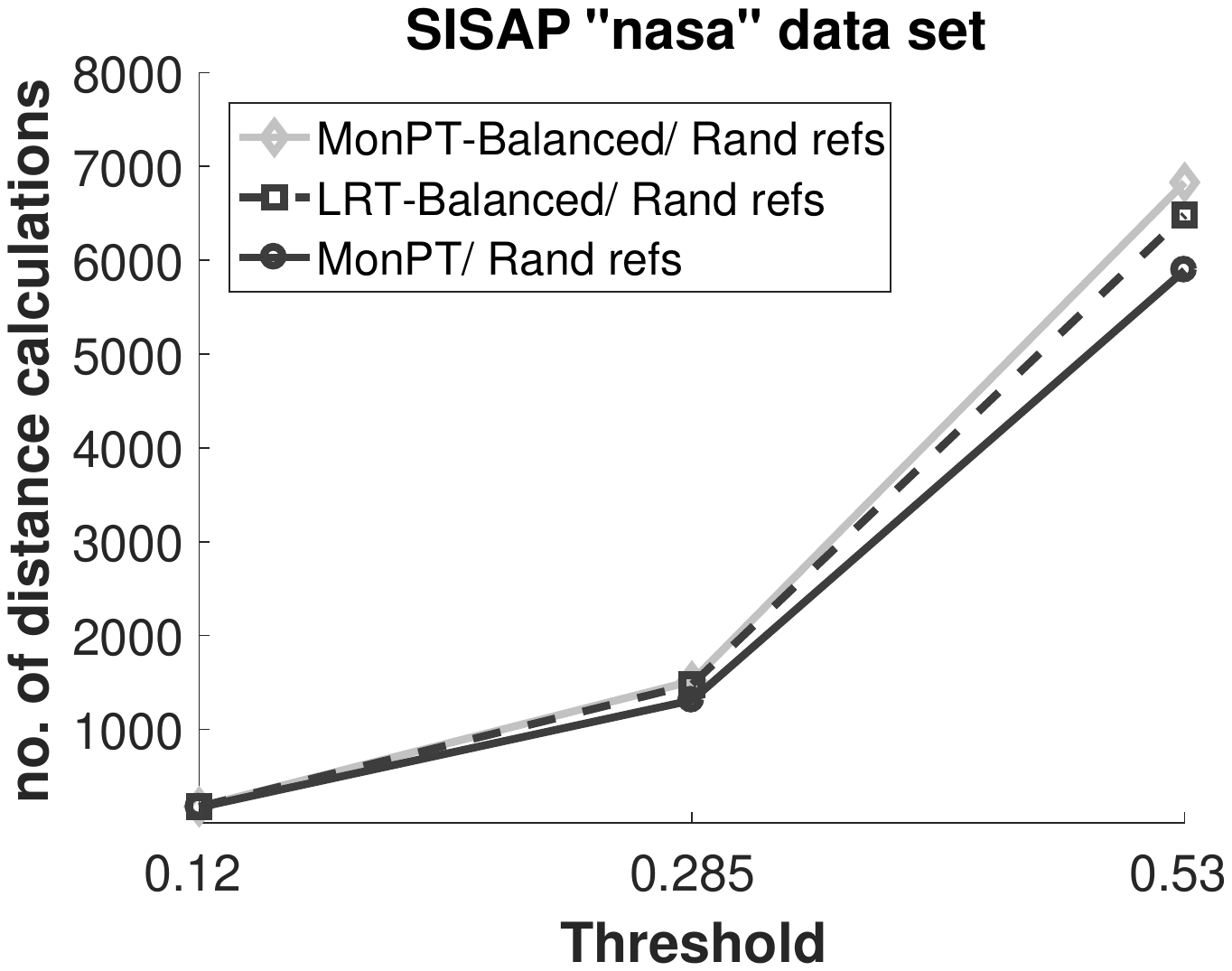} 
	\caption{Nasa --Monotone Tree (MonPT), Linear Regression Tree (LRT), and Balanced Monotone Tree with two different reference point selection strategies}\label{fig:LRTnasa}
\end{figure*}											

\section{Hyperplane Partition Indexes and the Four-Point Property}\label{sec:exp}

Having established the full generality of indexing in the supermetric domain, we now return to the Hilbert Exclusion principle and investigate its application over a range of hyperplane indexing structures. This is important in the light of the preceding discussion as, having established that issues such as balance and separation of reference points have quite different consequences, we need to understand the best indexing structure for taking advantage of the increased tractability. It is certainly not reasonable to assume  that the best indexing structures for metric spaces will also the the best for supermetric spaces.

The best recorded general performance for an exact-search partition-based indexing structure, before the identification of using the  four-point property within an exclusion mechanism, derives from the Distal Spatial Approximation Tree (DiSAT) \cite{dSatIS}. This is therefore the obvious comparison to make between using normal metric properties and the four-point  property over a space which has both properties; it allows the exactly same data structure to be measured with the different exclusion algorithms and in this sense is a very fair comparison. This comparison has been made in \cite{hilbert_embedding} and a significant improvement shown for using the four-point property: for the SISAP benchmark data sets, at the lower thresholds, typically around half the number of distance calculations are required when the four-point property is used over the same search index.

However, given the observations above on the different relative importance of the choice of reference point, it may be that the same data structure does not give the best  performance when used for a supermetric space; the main differentiation between previous versions of the Spatial Approximation Tree (SAT) index and the DiSAT is the choice of widely separated reference points at higher levels of the tree, and it is therefore possible that different optimising factors will occur within a supermetric space.

We therefore performed a thorough investigation on a number of different exact-search hyperplane tree structures, taking each possible orthogonal attribute separately and testing all possible combinations with both Hyperbolic and Hilbert Exclusion strategies to determine the best data structure for use in a supermetric space.

\subsection{Partition Trees}

\begin{algorithm}[tbp]
	{\small 
		\SetKwInOut{Input}{Input}
		\SetKwInOut{Output}{Output}
		\Input{Finite set of data objects $S\subseteq  U$}
		\Output{{Node of arity $n$: $ N=\left\langle \{p_1, \dots, p_n\}, \{N_1,\dots,N_n\}, \{cr_1, \dots, cr_n\} \right\rangle$  where $p_j \in S$, $N_j \in \text{Node}$, $cr_j\in \mathbb{R}$ for all $j$ from $1$ to $n$}}
		\smallskip
		\eIf{{$|S|\leq n$}}{{$N \gets \left\langle S,\emptyset, \emptyset \right\rangle$}\quad \texttt{//leaf node}}
		{select $\{p_1, \dots, p_n\}$ from $S$\;
			\ForEach{$p_i$ {in} $ \{p_1, \dots, p_n\}$}{
				$S_i \gets  \left\{s \in S \smallsetminus\{p_1, \dots, p_n\} |\, i = {\text{arg\,min}}_{{j=1,\dots,n}} d(s,p_j) \,\right \}$ \; 
				$cr_i \gets \min_{s \in S_i} d(s,p_i) $\;
				$\text{N}_x \gets$ CreateNode($S_i$)\;
			}
			{$N \gets \left\langle \{p_1, \dots, p_n\}, \{N_1,\dots,N_n\}, \{cr_1, \dots, cr_n\} \right\rangle$}
		}
	}
	\caption{{CreateNode}\label{alg_createnode}} 
\end{algorithm}
%

\begin{algorithm}[tbp]
	{\small 
		\SetKwInOut{Input}{Input}
		\SetKwInOut{Output}{Output}
		\Input{{$N=\left\langle \{p_1, \dots, p_n\}, \{N_1,\dots,N_n\}, \{cr_1, \dots, cr_n\} \right\rangle \in Node, q \in U, t \in \mathbb{R}$}}
		\Output{set $R= \{r \in S|\, d(q,r) \le t \})$}
		$R \leftarrow \emptyset$\;
		\ForEach{{$p_i$ in $\{p_1, \dots, p_n\}$}}{
			$d_i \gets d(q,p_i)$\;	
			\If{$ d_i \le t$}{
				$R \leftarrow R \cup \{p_i\}$
			}
		}
		\If{$\{N_1,\dots,N_n\} \neq \emptyset$}{
			$ Excs \gets \emptyset$\;
			\ForEach{$p_i$ in $\{p_1, \dots, p_n\}$}
			{
				\If{$d(q,p_i)  \ge cr_i + t$} {$Excs\leftarrow Excs \cup \{i\}$}
				\For{{$p_j \in \{p_1, \dots, p_n\}\smallsetminus{p_i}$}}{\If{$d(q,p_i) - d(q,p_j) > 2t$}{$Excs\leftarrow Excs \cup \{i\}$}}
			}
			\ForEach{$i$ in $\{1,\dots,n\}\smallsetminus Excs$}{
				{$R \leftarrow R \cup \text{QueryNode}(N_i)$}
		}}
		
	}
	\caption{{QueryNode} \label{alg_query_node}}
\end{algorithm}

The basic structure of a partition tree is a recursively defined, n-ary tree. Each child of a parent node is governed by a single reference point, and  every element of the  data set contained below any parent node is associated with the child node whose reference point is closest to that element. The basic construction of a partition tree is given in Algorithm \ref{alg_createnode}.

Query of such a tree allows a number of exclusion possibilities. Most simply, the distance from the query to each reference point is  calculated; any partition may be excluded if this distance is greater than the cover radius ($cr_i$) added to the query threshold. In addition, for every partition, it may be excluded from the search based on the relative distances of the query to every other reference point, using the hyperplane exclusion principle. The basic query algorithm, not using the four-point property, is given as Algorithm \ref{alg_query_node}. This can be changed to take advantage of the four-point property by the replacement of the inner partition test {(lines \texttt{14-18})}:

\begin{align*}
&d(q,p_{i}) - d(q,p_{ j}) > 2t	\\
\text{is replaced by}	\\
&\frac{d(q,p_{ i})^2 - d(q,p_{ j})^2}{d(p_{ i},p_{ j})} > 2t	\\
\end{align*}

The extra required term, \h{$d(p_{ i},p_{ j})$} can be calculated at build time and stored for a small extra space  cost, explained in Section \ref{section_hilbert_costs}.

The many different types of tree we tested are differentiated by how the reference set is chosen at each tree node. We tested a number of variants of  trees according to the following largely  orthogonal principles:

\begin{description}
	\item[Pure SAT property] Each node of a purely-formed Spatial Approximation Tree has the property that no values from within the data set are closer to the ``centre" node (the reference point in the parent node associated with each child node) than they are to any of the reference points one level down within the tree. This principle of construction allows further exclusion possibilities, as during a query the maximum distance between the query and any higher-level reference points may be passed recursively; if this distance is greater than $\max(d(q,p_j)), i \neq j$ then it may be used to attempt to exclude $N_i$ from the search.
	
	Any serial selection of reference points   requires, as each node is constructed, that any point closer to the centre node than any previously selected reference point is added to the set of reference points. As pointed out by the authors of  \cite{dSatIS} the construction has very different properties depending on the order in which the contained set is considered.
	
	Two variants of such ``pure" SATs were tested; for \texttt{\small sat\_pure},  we considered the data set of inclusion in the reference set in order of distance from the centre node, and for \texttt{\small sat\_distal\_pure} we considered them in reverse order. We also tried some hybrids but did not discover any interesting results, so these are not reported.
	
	For using the Hilbert Exclusion mechanism for trees with this pure SAT property, the distances to ancestor nodes can still be used, but (i) all distances need to be recursively passed, rather than just the maximum, and (ii) at build time, all distances from each $p_i$ to all ancestors also require to be calculated and stored. This is because the Hilbert exclusion condition requires all three distances among any two reference points and the query to be known. We tested both construction and query for any extra cost associated with this extra information flow, and it was found to be  insignificant.
	
	\item[SAT construction]
	Faster versions of SAT as reported in \cite{dSatIS} do not maintain the pure SAT property, but instead reuse the core reference point selection algorithm (traversing $S$ according to an imposed order and adding reference points whenever they are closer to the centre node than any existing reference point), however  this process is terminated according to the number of reference points selected. This is because the pure SAT algorithm, applied to the distal ordering of values from the centre, leads to very wide, shallow trees which do not lead to good performance when using Hyperbolic exclusion. The extra exclusion possibilities from using the parent reference point distances are lost, but wider separation of reference points was found to lead to more exclusions. 
	
	Therefore the extra factor of maximum branching factor is considered. We considered two; \h{\texttt{\small sat\_distal\_fixed} uses} a fixed value of 4, found in \cite{dSatIS} to give good performance, and \h{\texttt{\small sat\_distal\_log} uses} a dynamic value selected according to the data size, chosen not to exceed the natural logarithm of the data size, thus reducing as the tree is descended.

	Choice of the centre point for the head node, for best performance, is described in \cite{dSatIS} as being acheived by choice of an outlier, this giving the $\text{SAT}_{out}$ class of algorithm\h{ (\texttt{\small sat\_global\_fixed} and \texttt{\small sat\_global\_log})}; we confirmed this result and therefore reused this strategy for all of our experiments.
	
	Finally, the $\text{SAT}_{glob}$ class of tree \h{uses} 
	a single ordering for consideration of reference point selection, based on an ordering of the whole data set from the centre node of the entire tree, rather than the centre of each node.

	\item[Non-SAT construction]
	Finally, we considered a number of partition trees \h{(\texttt{\small hpt\_*})} where the choice of reference point was made independently of their distances from the parent reference point. Three arities were chosen; \textit{fixed} and \textit{logarithmic} as above, but also a \textit{binary} version was constructed. Two strategies for reference point selection were used to fill these arities: \textit{random} selection, and one using  the \textit{FFT} algorithm \cite{Dasgupta2005,Gonzalez1985}.
\end{description}

\subsection{Experimental Procedure}
The above classification leads to the following set of tree structures used for experiments:
{\small\texttt{sat\_pure, sat\_distal\_pure, sat\_distal\_fixed, sat\_distal\_log,\\ sat\_global\_fixed, sat\_global\_log
		, hpt\_fft\_binary, hpt\_fft\_fixed, hpt\_fft\_log, hpt\_random\_binary, hpt\_random\_fixed, hpt\_random\_log} }
-- each of these should have a clear meaning given the above description. For each tree, both Hyperbolic and Hilbert exclusion mechanisms were tested, leading to a total of 24 different search indexes.

Each was tested against the SISAP benchmark data sets \emph{colors} and \emph{nasa}~\cite{SISAP_man} using Euclidean distance. For each test, three different query thresholds were used as is standard for the SISAP benchmark sets.  10\% of the data was randomly removed to act as a query set. 
\begin{table*}[tb]											
	\renewcommand{\arraystretch}{1.2}											
	\begin{center}											
		\small{\begin{tabular}											
				{											
					|l|											
					l|											
					l|
					l|	
					l|										
				}
				\hline											
				\textbf{Data Set}		&	\textbf{\# elements}	&	\textbf{feature}	&	\textbf{dim}	&	\textbf{dist}	\\
				\hline
				SISAP Colors~\cite{SISAP_man}	&	112,682	&	Color Histograms	&	112	&	$\ell_2$	\\
				SISAP Nasa~\cite{SISAP_man,nasaDataset}	&	40,150	& PCA-red. Color Histograms	&	20	&	$\ell_2$	\\ 
				\hline
		\end{tabular}	}										
	\end{center}											
	\caption{Data Sets Statistics}\label{tab:datasets}											
\end{table*}

Different distance metrics with the four-point property were also tested but showed no significantly different results, so we report only Euclidean. Actual threshold values used   are as shown in Table \ref{table_sisap_thresh}.

\begin{table}[tbp]
	\begin{center}
		\begin{tabular}{|c|c|c|c|}
			\hline
			\textbf{Data Se}t&$\mathbf{t_0}$ \h{(0.01\%)}&$\mathbf{t_1}$ \h{(0.1\%)}&$\mathbf{t_2}$ \h{(1\%)}\\
			\hline
			nasa & 0.120 & 0.285 & 0.530\\
			colors & 0.052 & 0.083 & 0.131\\
			\hline
		\end{tabular}
	\end{center}
	\caption{\h{Experimental threshold values that return around
			0.01\%, 0.1\% and 1\% of the data sets}}
	\label{table_sisap_thresh}
\end{table}%

For sake of space, we give only the main results: for each index, and each data set, we give the number of distances per query at a single query threshold.

We also tested query times; in all cases the number of distances was directly proportional to the query time, which is not surprising as all of the data sets used fit comfortably within the main memory and this is by far the dominant cost of the query. It is useful to confirm however that the extra administrative cost associated with the Hilbert exclusion is negligible with respect to distance costs.

All tests were executed in Java, using the same   Java abstract tree construction and query classes, specialised  only according to the reference point selection strategy and query-time exclusion strategy. As a final semantic check, all results were cross-checked against a serial (exhaustive) search to ensure  consistency and therefore correctness. All code used is available in a public 
repository\footnote{https://bitbucket.org/richardconnor/metric-space-framework.git}.

For each test, multiple tree builds were performed and mean values are presented.  For each build, the data was presented in randomised order, as the order of selection during tree build can have a significant serendipitous effect on performance. Tests were repeated until the standard error of the mean was $\le 1\%$, which implies that all of the differences reported are statistically significant.

Results are shown in Figure \ref{fig_trees_colors} and  \ref{fig_trees_nasa}.

\begin{figure}[tbp]
	\centering
	{\includegraphics[trim=30mm 90mm 40mm 90mm,width=0.7\columnwidth]{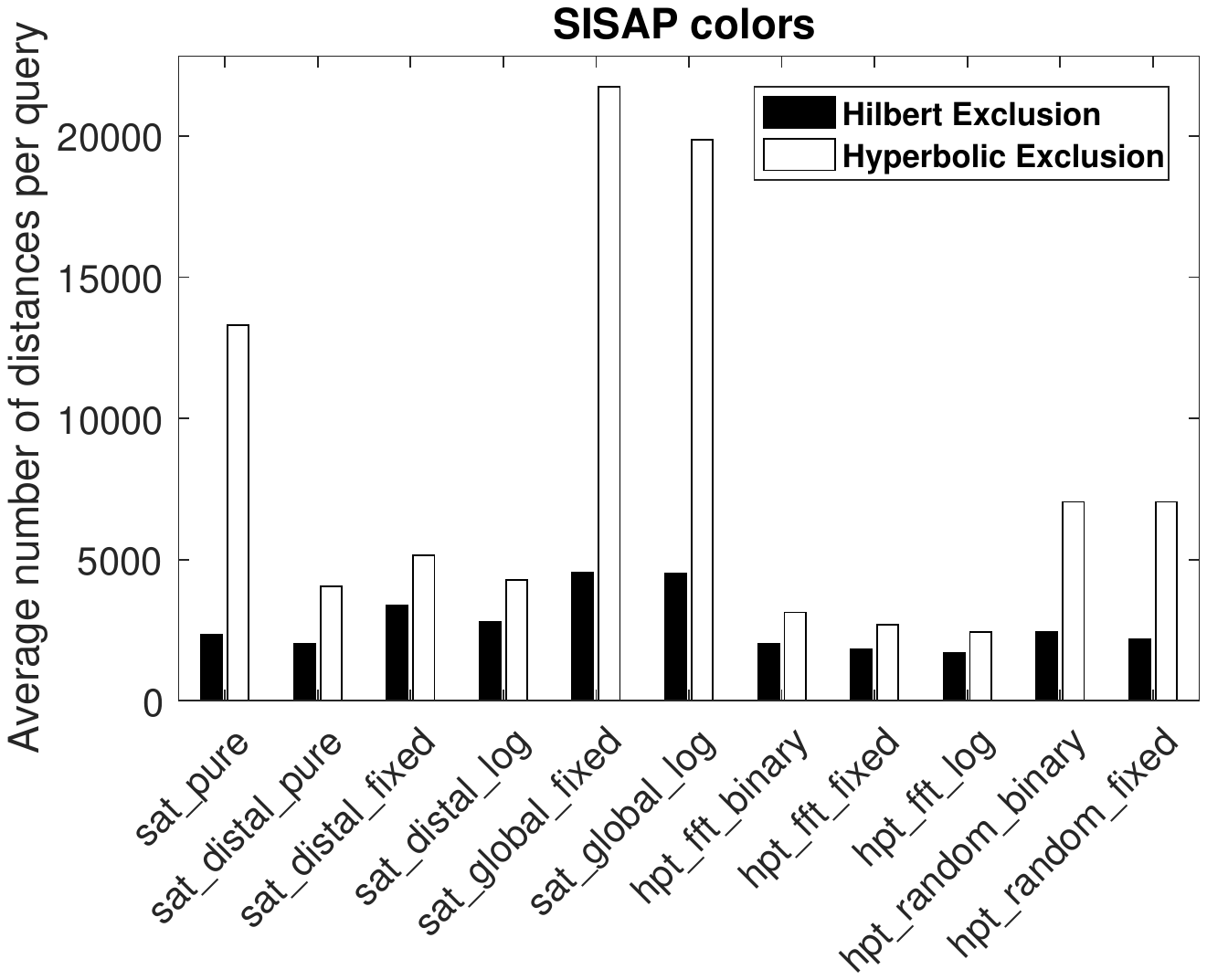}}
	\caption{Partition Trees: SISAP \emph{colors} at threshold $t_0$}
	\label{fig_trees_colors}
\end{figure}
\begin{figure}[tbp]
	\centering
	{\includegraphics[trim=30mm 90mm 40mm 90mm,width=0.7\columnwidth]{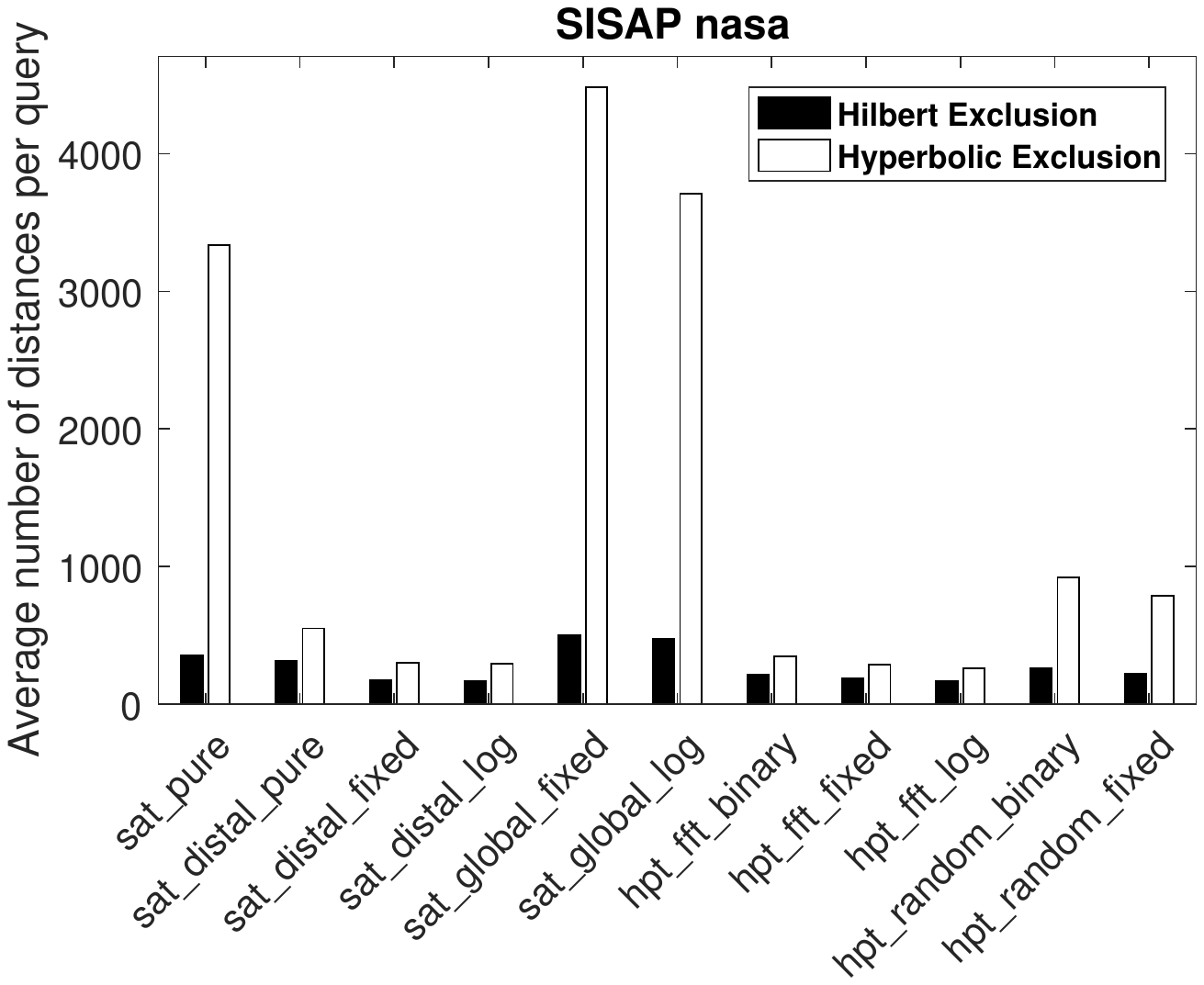}}
	\caption{Partition Trees: SISAP  \emph{nasa} at threshold $t_0$}
	\label{fig_trees_nasa}
\end{figure}

\subsection{Analysis}
The most obvious conclusion is that the supermetric exclusion always gives better performance; while this is actually a guarantee as shown in \cite{hilbert_embedding} the interesting point is the magnitude of the improvement, which in some cases is quite startling.

Although not quite the absolute best performance, the greatest improvement is in the pure SAT indexes, both classic and distal variants which in fact seem to give around the same performance. This is worthy of further study; as mentioned, the shape of these trees is very different, the classic SAT giving a relatively small branching factor against a very large branching factor at the higher levels of the distal SAT. The inventors of the distal SAT compromised the SAT property early on, presumably because performance was badly affected  by this property. Using the four-point property seems to overcome this. As these data sets are relatively small there is no real advantage to a shallow tree, but this may well be different with very large data.

The lack of variance for the  four-point exclusion across all the different structures is also notable; this confirms our earlier hypothesis that the actual exclusion power of the Hilbert mechanism is much less affected by the choice of reference point, and certainly confirms that putting  huge computational resources into building expensive data structures may be far less worthwhile in this context.

Finally, we note the best performance data structure considered here, the log-sized hyperplane tree using the FFT algorithm to choose reference points. Paradoxically, this is one of the simplest, and fastest, structures to construct. It is likely that using a more sophisticated cluster-finding algorithm such as k-means or k-medoids may perform a little better, although at much higher tree build cost; given the rather small incremental improvement however of FFT over random, we are not convinced that this would be worthwhile in many cases.

And as a last word: we note that the  values of 1,704 distance measurements per query achieved over the SISAP \emph{colors} data set, and 171 measurements per query over the \emph{nasa} data set are, for the moment, new performance records against this benchmark.

\section{The  Cost of Hilbert Exclusion}
\label{section_hilbert_costs}
While it has been shown that Hilbert Exclusion performs better than Hyperbolic in run-time cost, there is an extra space cost involved as more information is required: the distance between the reference points is required for the query-time exclusion calculations, where it is not for the hyperbolic exclusion calculation. It is thus important to assess the extra overhead in time and space.

In all cases, the choice of reference points is made during the building of the index structure; therefore it is possible for the distances to be calculated at build time and stored. Feasibly, they could instead be calculated at query time if the storage overhead was relatively great compared with the extra query-time cost; however here we show that it is not.

In the case of binary trees, the extra space overhead is a single distance value, or 4 bytes\footnote{
	as the value is only used for additive arithmetic and is not critical for correctness, single precision is sufficiently accurate
}, per node. Even the leanest tree implementation will have a per-node space overhead much greater than this, although of course this depends on the language and implementation tactics used. A modern JVM has a minimum object overhead of 16 bytes, along with a further 32 bytes to store pointers to two subtrees even before any other node information is considered, such as object ids for the reference objects, cover radii, etc. Realistically the extra overhead is likely to be much less than 10\%; given the further invariant that the number of internal tree nodes is less than one-half the number of data\footnote{
	for a monotone tree;  less  for other types
}, and each data object is likely to be much bigger than a tree node, the space overhead is minimal and unlikely to be significant in any realistic scenario.

For general hyperplane trees with more than two reference points the overhead is potentially  greater as all inter-reference point distances are required, giving a theoretical $\mathcal{O}(n^2)$ space cost. Pragmatically however the values of $n$ involved are fairly small; furthermore as we are dealing with proper distances we only need store the upper triangular matrix, so the space overhead is $n \choose 2$ rather than $n^2$. For example, in \cite{dSatIS} the authors observe that a DiSAT branching factor of 4 gives optimal performance in some contexts; here we can replace the $\mathcal{O}(n^2)$ observation with a constant value of 6, i.e. 24 bytes per tree node. In the case of a quadtree, the number of internal tree nodes will be less than approximately a third of the data size, giving a maximum overhead of less than 8 bytes per data object

Finally we consider the log-sized node strategy, where the number of reference points at each tree node is approximately the log of the volume of data stored below the node. This  leads to much bigger overhead at the root of the tree; for example the root node of a tree for $10^{10}$ data objects has 23 reference points, requiring 1KByte of overhead. However these   trees are correspondingly shallower, and the node size decreases rapidly as the tree extends downwards. There is a  recurrence relation to estimate the space overhead in bytes for a balanced tree:
\begin{align*}
\textit{overhead} (N) \quad&=\quad 0, \qquad N \le 2\\
\textit{overhead} (N) \quad&=\quad {|p| \choose 2} \times 4 +
|p| \times \textit{overhead}(\tfrac{N - |p|}{|p| })\\
\text{where}\quad |p| \quad&=\quad \lfloor \log N \rfloor
\end{align*}
When applied to  any large size the overhead turns out to be only around  one byte per data object.

There is  also clearly an extra run-time cost in construction, from the measurement of extra distances. This is of much less concern; in any partition strategy, for every node built, there is a requirement to measure the distances between every data item below the level of that node against all of the reference points. Without further analysis it seems clear that the extra overhead of measuring the distances among the nodes is relatively trivial.

Experimental evidence supports the notion that  the extra overhead, in both time and space, is insignificant, independent of the strategy used. We have tried to detect significant differences in either construction time or space in the large experiments described in Section \ref{section_large_experiments}, but in all cases the differences have been hidden in the noise caused by the introduction of randomisation to the construction process.

\section{Spaces with Larger Scale and Higher Dimensionality}
\label{section_large_experiments}
In this section we investigate extending the  use of the  supermetric property to larger and higher-dimensional data sets. The purpose is to explore  how the behaviour of the mechanisms, which have been shown to give good results over (relatively) small benchmark sets,  alters with sets that are inherently more difficult to index.

We perform three types of test to demonstrate these behaviours:
\begin{description}
	\item[ ] 
\end{description}

\begin{description}
	\item[increasing dimensionality] 
	
	In these tests, we generate evenly-spaced points within generated Euclidean spaces of increasing dimension and test Hyperbolic and Hilbert exclusion mechanisms for query performance. These tests show the effect that increasing dimensionality has on the relative performance; pragmatically we show that at the point where increasing dimensionality starts to make search intractable, the use of Hilbert exclusion gives an extra 2-3 dimensions for the same level of performance.
	
	\item [``real-world" high-dimensional data] 
	
	In these tests we use GIST representations of the MIR-Flickr \cite{mirflickr} data set of one million images to perform a near-duplicate search; these are large data comprising 480 floating-point numbers, tested with various metrics. These tests show that the efficacy of 
	\h{Hilbert exclusion} does not seem to be affected by the choice of metric.

	\item [increasingly large data sets] 
	
	In these tests we use 80 dimensional  MPEG-7 Edge Histogram data taken from the CoPhIR \cite{CoPhIR} images set . Tests are made over increasingly large subsets (between 1 and 16 million images) to show how the mechanisms scale as the data size increases.
\end{description}

\subsection{Increasing Dimensions}
For each dimension between 2 and 20 inclusive, evenly distributed Cartesian points were generated within the unit hypercube. For each dimension, a Euclidean space of one million data points was generated, and one thousand threshold queries were executed over each space. At each dimension a threshold was selected with a radius calculated to give one-millionth of the volume of the unit hypercube\footnote{
	For dimension $n$, radius $r_n = \frac{\Gamma(\tfrac{n}{2} + 1)}{\pi^{\tfrac{n}{2}}}$, where $\Gamma$ is Euler's gamma function
}.

After first confirming that the log-sized hyperplane partition tree is still the most efficient index at all dimensions, measurements were made for four variants to establish the added value of the Hilbert exclusion mechanism. Trees were built using both random and FFT selection strategies for each node, and for both of these variants, querying was performed with, and without, the Hilbert exclusion mechanism. As previously noted, the same instance of the built data structure can be used in either way, as long as the metric has the four-point property. The figure recorded is the  number of distance calculations required per query over the data set; at each dimension, each experiment was repeated until the standard error of the mean was less than 1\%.

\begin{figure*}[tbp]
	\begin{center}
		{\includegraphics[trim=15mm 20mm 15mm 20mm,clip,width=0.65\textwidth]{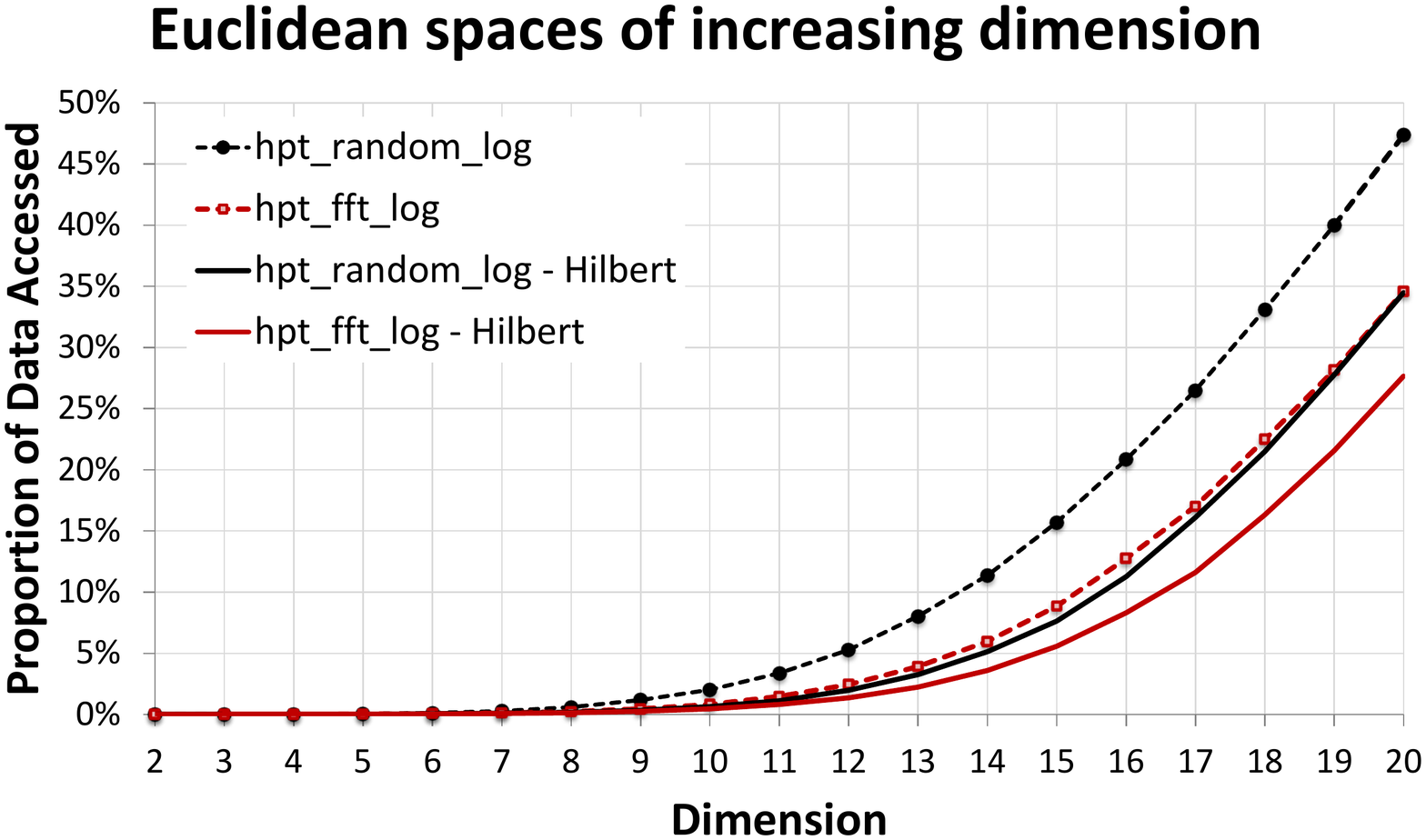}}\\%
		{\includegraphics[trim=15mm 20mm 15mm 20mm,clip,width=0.65\textwidth]{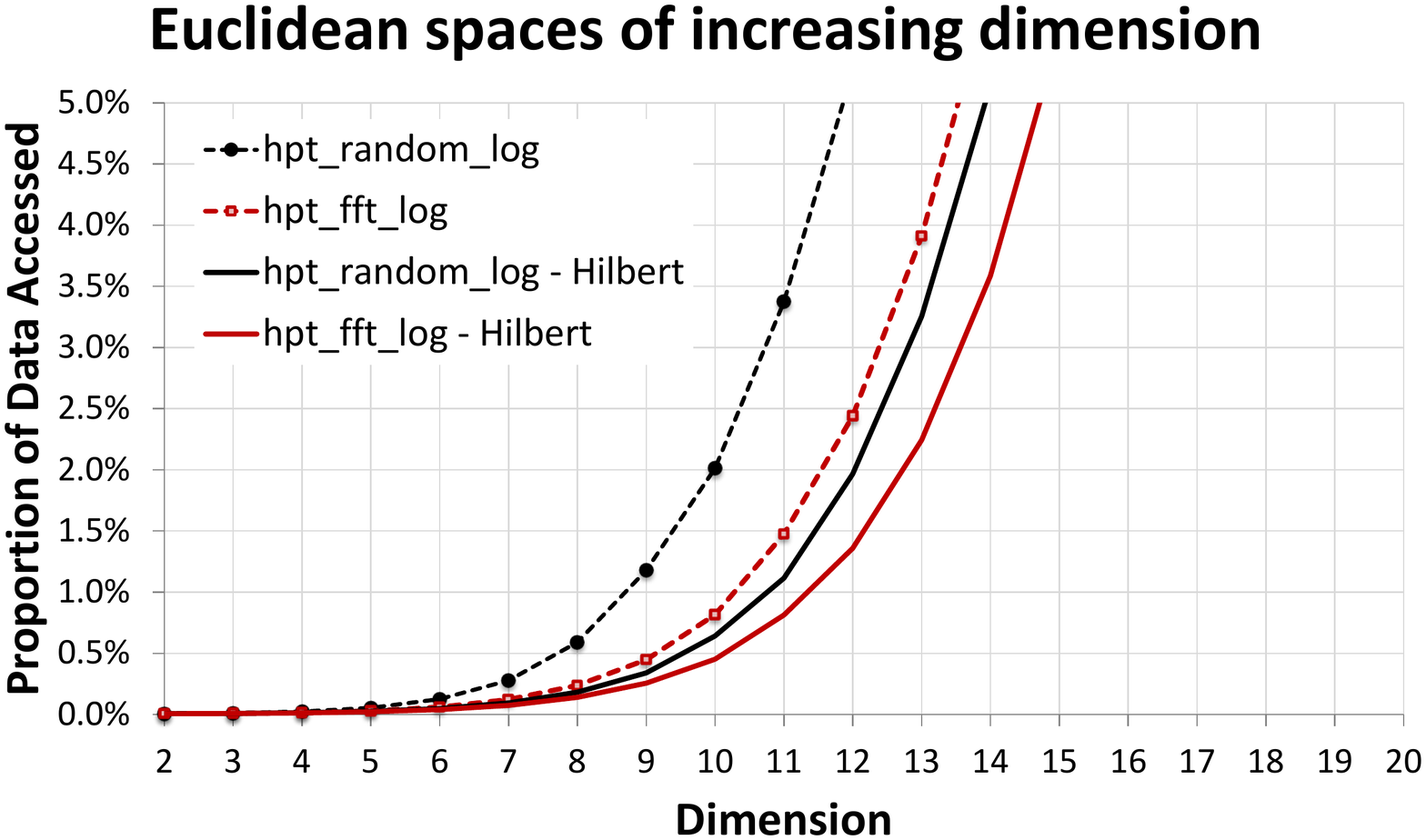}}%
		\caption{
			The effect of increasing dimensions on the four HPT mechanisms. The right-hand side is a magnified version of the left, allowing the very significant performance differences at lower dimensions to be seen better.
		}
		\label{figure_inc_dims}
	\end{center}
\end{figure*}

Figure \ref{figure_inc_dims} shows the outcomes. It can be seen that, across all dimensions, the FFT variant is better than a random choice of reference points for either exclusion mechanism, and more significantly for our purposes that the Hilbert exclusion variant is substantially better than the Hyperbolic.


Most importantly perhaps is the observation that these two improvements are almost orthogonal, and between around 8 and 12 dimensions the result is a four-fold increase in performance.

The tree \texttt{hpt\_fft\_log} (FFT pivot selection) using the Hilbert exclusion mechanism shows quite similar performance to the tree \texttt{hpt\_random\_log} (random pivot selection) using the Hyperbolic exclusion mechanism; although subject to heuristics and uncertaintly, in our experiments the FFT-based choice of reference points gives a  build cost of around five times that of the  hyperplane tree with randomly selected reference points, and indeed the latter is one of the cheapest indexing mechanisms to build; if build time is an important consideration, this could give the best compromise.

An alternative view of the results is to consider where the different lines cross a given horizontal boundary in the chart. For example, if a particular situation indicates that accessing no more than 2.5\% of the data is required to give sufficient performance, than this can be achieved with a data set whose dimensionality is around 13 using FFT and Hilbert exclusion, whereas only around 10 can be achieved, from the same indexing mechanism,  without these.

\subsection{MirFlickr/GIST and Near-Duplicate Detection}

\begin{table*}[tb]											
	\renewcommand{\arraystretch}{1.2}											
	\begin{center}											
		\small{\begin{tabular}											
				{											
					|l|											
					l|											
					l|											
					l
					l	
					l	
					l
					l|									
				}
				\hline											
				\textbf{Metric}		&	\textbf{IDIM}	&	\textbf{Unit cost}&\textbf{$t_1$}	&	\textbf{$t_2$}	&	\textbf{$t_3$}&	\textbf{$t_4$}	&	\textbf{$t_5$}	\\
				\hline
				Euclidean	&	14.3 &0.0016		&0.016&0.044&0.070&0.096&0.125	\\
				Cosine		&	15.5	& 0.0013	& 0.006& 0.016& 0.027&0.037& 0.048	\\
				Jensen-Shannon & 10.8	 &0.017	&0.005&0.011&0.018&0.024&0.032	\\
				Triangular		& 12.1	&	0.0029	&0.006&0.013&0.021&0.029&0.037	\\ 
				\hline
		\end{tabular}	}										
	\end{center}											
	\caption{GIST: IDIM, cost per distance measurement, and thresholds used}\label{tab:gist_stats}											
\end{table*}

\begin{figure*}[tbp]
	\begin{center}
		{\includegraphics[trim=15mm 20mm 15mm 20mm,clip,width=0.65\textwidth]{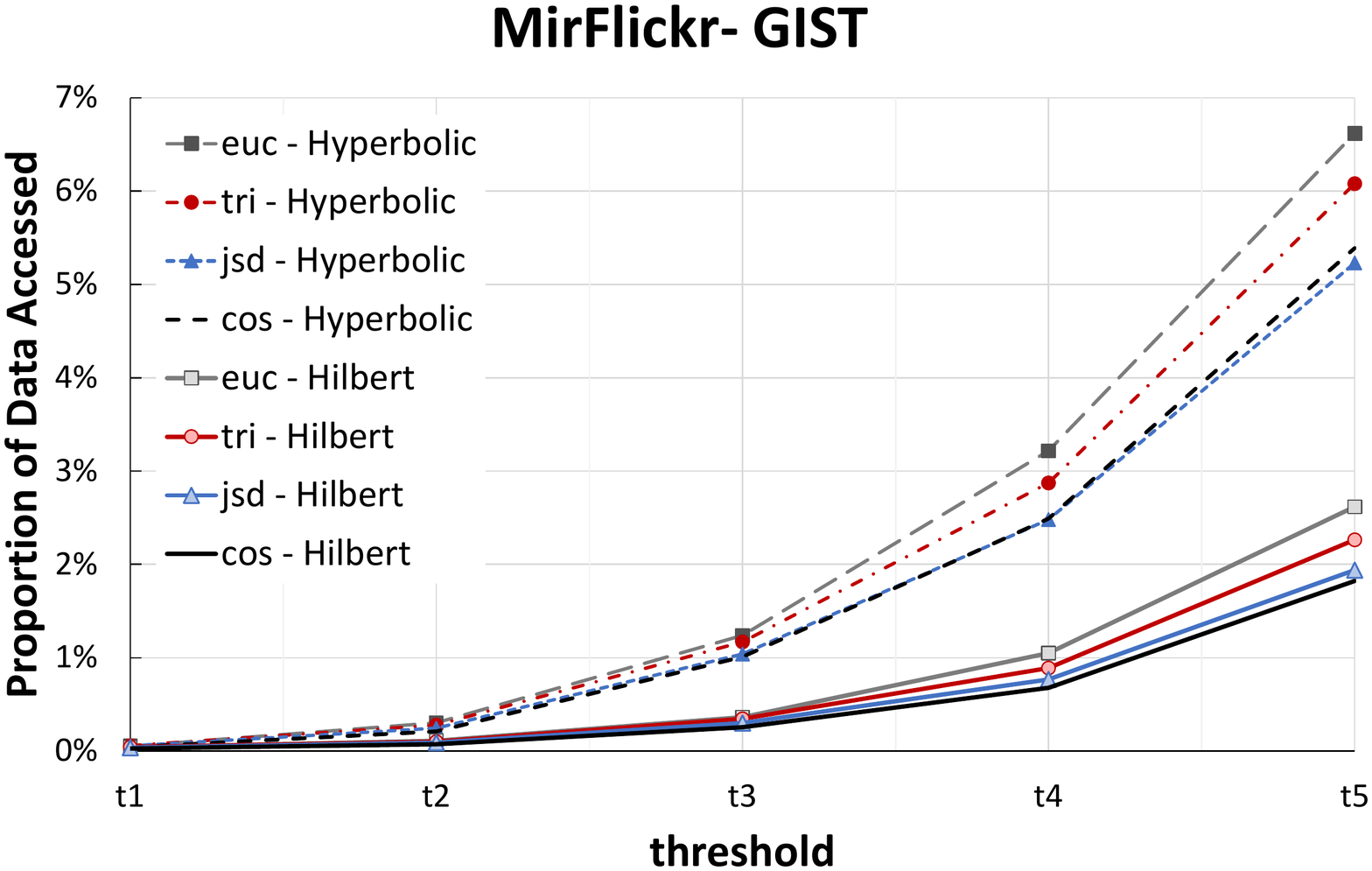}}%
		\\
		{\includegraphics[trim=15mm 20mm 15mm 20mm,clip,width=0.65\textwidth]{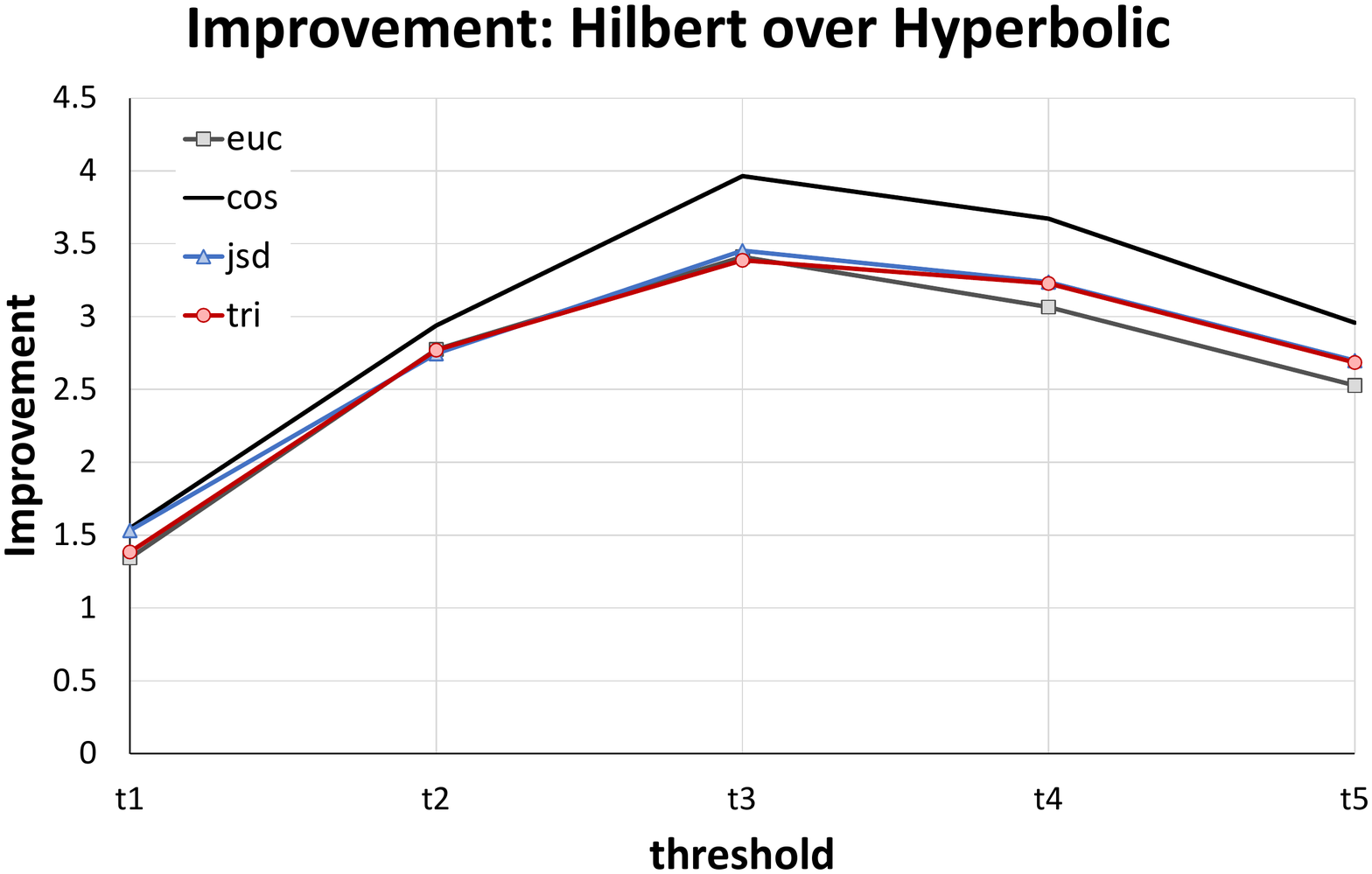}}%
		\caption{
			Four different metrics over GIST data at different thresholds, with and without using Hilbert exclusion.
		}
		\label{figure_gist}
	\end{center}
\end{figure*}
In these experiments we test a large data set for a real-world purpose, namely the detection of near-duplicate images. In previous work we have shown the use of the GIST characterisation gives the best tractable test for near-duplicate image detection within large sets of images \cite{connor2016quantifying};  these tests are inherently expensive because of the data size, and efficient similarity search is  important to give tractability. Each GIST object\footnote{
	using GIST parameters: 4 windows, 6 scales and 5 orientations per scale, taken from a monochrome 255 $\times$ 255 image with no border
}  comprises 480 dimensions of floating point numbers, which using  IEEE single-precision format gives an object size of almost 2KBytes per object, i.e. just under 2GBytes per million images.

While the intrinsic dimensionality of these spaces is relatively high -- around 10-15 depending on the metric -- the required search thresholds are quite low, therefore giving a nice example of spaces where metric search is particularly appropriate for the task in hand.

We used the Mir-Flickr \cite{mirflickr} set of one million images and generated GIST representations. We have previously demonstrated the use of this collection as a benchmark for near-duplicate image detection; the collection by chance contains around 2,000 clusters of near-duplicate images which we have identified, allowing both sensitivity and specificity to be accurately tested for different metrics and thresholds \cite{connor:mfnd}. The GIST representations can be used for this purpose with any of Euclidean, Cosine, or Jensen-Shannon distances. A key aspect of such classification functions with very large collections is their specificity, which must be high to avoid very large numbers of false positives. All of these tests maintain high specificity up to a sensitivity of around 50\%; to test the efficiency gains of the Hilbert Exclusion we therefore tested searches over the collection at five different thresholds, representing for each metric sensitivity of 10\% to 50\%, after which point a  fast drop-off in specificity occurs. Table \ref{tab:gist_stats} gives, for each metric, the intrinsic dimensionality, the mean cost per distance measurement in milliseconds, and the thresholds used to search.

For each of the three metrics, the first 1,000 images were used as queries against the remainder of the data. A log-sized hyperplane partition tree was constructed, with reference points chosen using the FFT technique. These trees were then tested with and without use of the Hilbert exclusion; in each test the number of distance calculations and query time were noted. These were found to be almost exactly directly proportional and so only the number of distance calculations are presented, given as the mean proportion of the total data size tested per query.

Figure \ref{figure_gist} shows results from these experiments. It appears that the advantage given by using the Hilbert exclusion mechanism is relatively independent of the metric being considered. In all cases it is highly significant, giving a performance improvement of 2.5 to 3 times for all metrics, even at the top end of the thresholds tested.
\subsection{Increasing Scale}
In these experiments, we investigate how  the advantages shown by the Hilbert property are affected by the scale of the data. To measure this, we used  MPEG-7 Edge Histogram descriptors extracted from the CoPhIR \cite{CoPhIR} image data set, using the first 80 dimensions
of the raw data. We queried increasingly large subsets of the data, ranging from one million to sixteen million images, to test the scalability of the different search mechanisms. Results are reported for Euclidean distance; we repeated the tests using other metrics and found no interesting differences.

We sampled $10^9$ randomly selected distances to measure IDIM and to choose search thresholds; the IDIM of the data was measured as 7.5, and three thresholds were selected to return $10^{-8}$, $10^{-7}$ and  $10^{-6}$ of the data per query\footnote{
	0.0196, 0.0834 and 0.1815 respectively
}; these thresholds are small but this is appropriate as data becomes larger.

With relatively smaller threshold and larger data we did not make assumptions about which mechanisms of those tested earlier would perform best; we tried them all, and report here the most interesting representative results. As there is some anecdotal evidence that single-pivot strategies can be more effective that hyperplane partitions as thresholds decrease, we also included a vantage point tree \cite{Yianilos1993} in the tests.  For each search structure, we performed 1,000 queries selected randomly from a different part of the set and measured the number of distance calculations performed; we present these as a proportion of the data access per query.

Figure \ref{figure_scaling} shows the outcomes. We present results for: log-sized hyperplane trees, with Hilbert and Hyperbolic exclusion; monotone (binary, unbalanced) hyperplane trees, again with Hilbert and Hyperbolic exclusion; a balanced vantage point tree, and a  linear regression tree. Reference points for the log-sized hyperplane trees are selected using FFT, and for the monotone binary trees by simply selecting the furthest object from the inherited reference point.

For all mechanisms and thresholds, it can be seen that as the size of the dataset increases, the proportion of data accessed decreases. The rate of this decrease demonstrates scalability of the mechanism. The value of the Hilbert exclusion is very marked, especially with the larger threshold values. It is interesting to note that the vantage point tree performs very well with a very small threshold, but is relatively much worse as the thresholds get larger. Although a less marked effect, the log-sized hyperplane tree appears to scale slightly better than the binary version. Finally, we note the best overall performance achieved by the linear regression tree; this version is an early attempt at using the extra flexibility allowed by the stronger geometry of the supermetric space, and demands further research.

\begin{figure*}[tbp]
	\begin{center}
		{\includegraphics[trim=15mm 20mm 15mm 20mm,clip,width=0.75\textwidth]{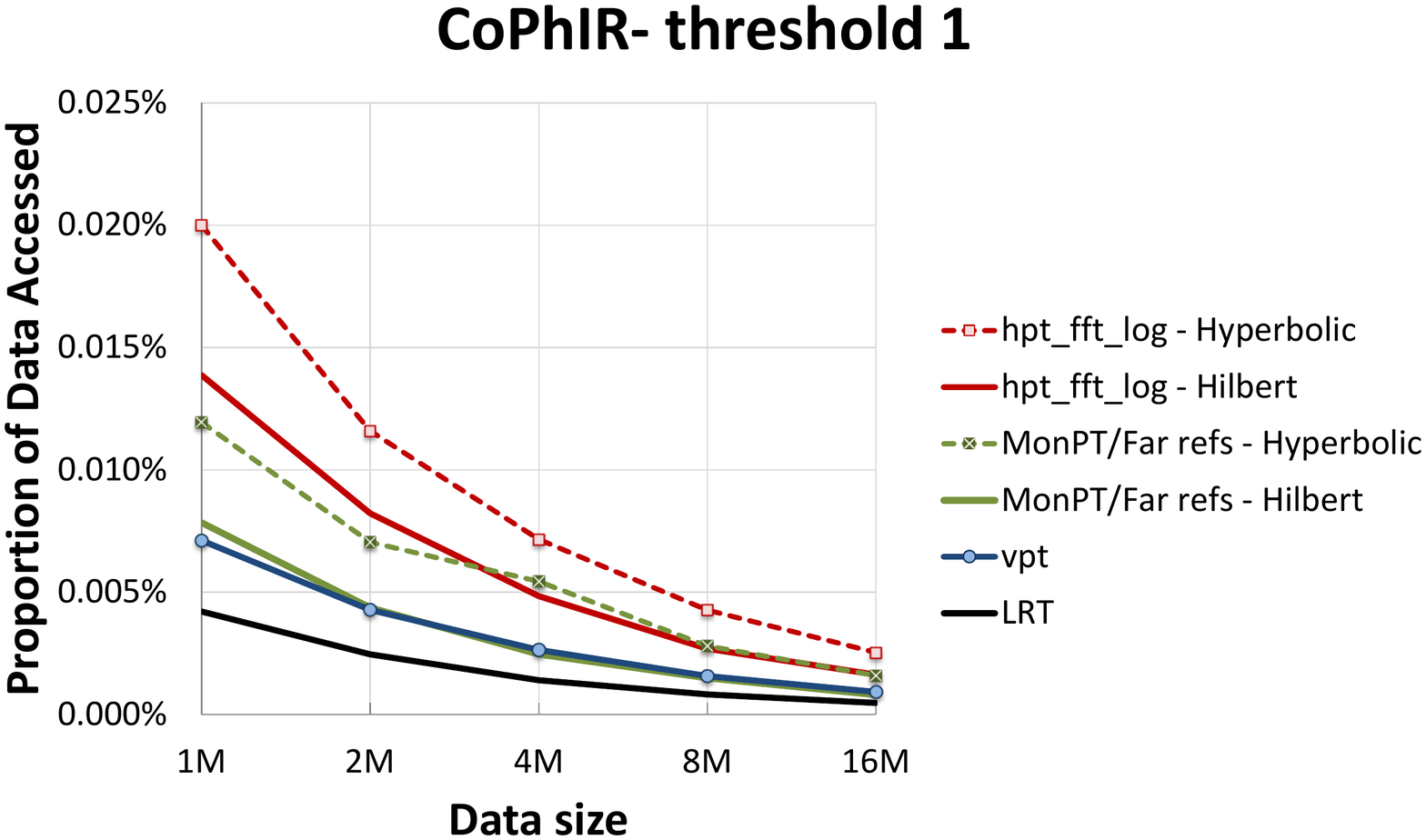}}\\%
		{\includegraphics[trim=15mm 20mm 15mm 20mm,clip,width=0.75\textwidth]{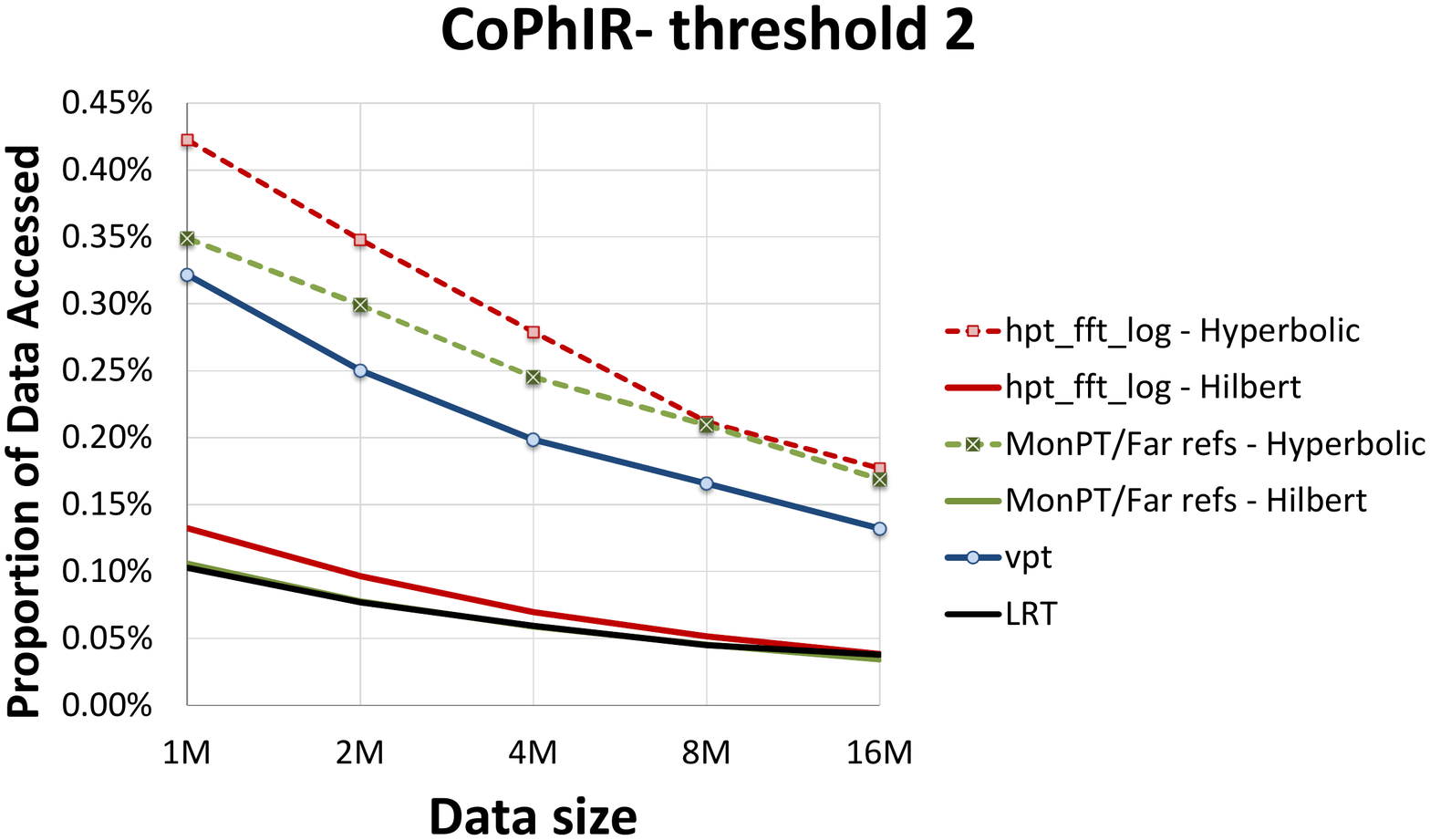}}\\%
		{\includegraphics[trim=15mm 20mm 15mm 20mm,clip,width=0.75\textwidth]{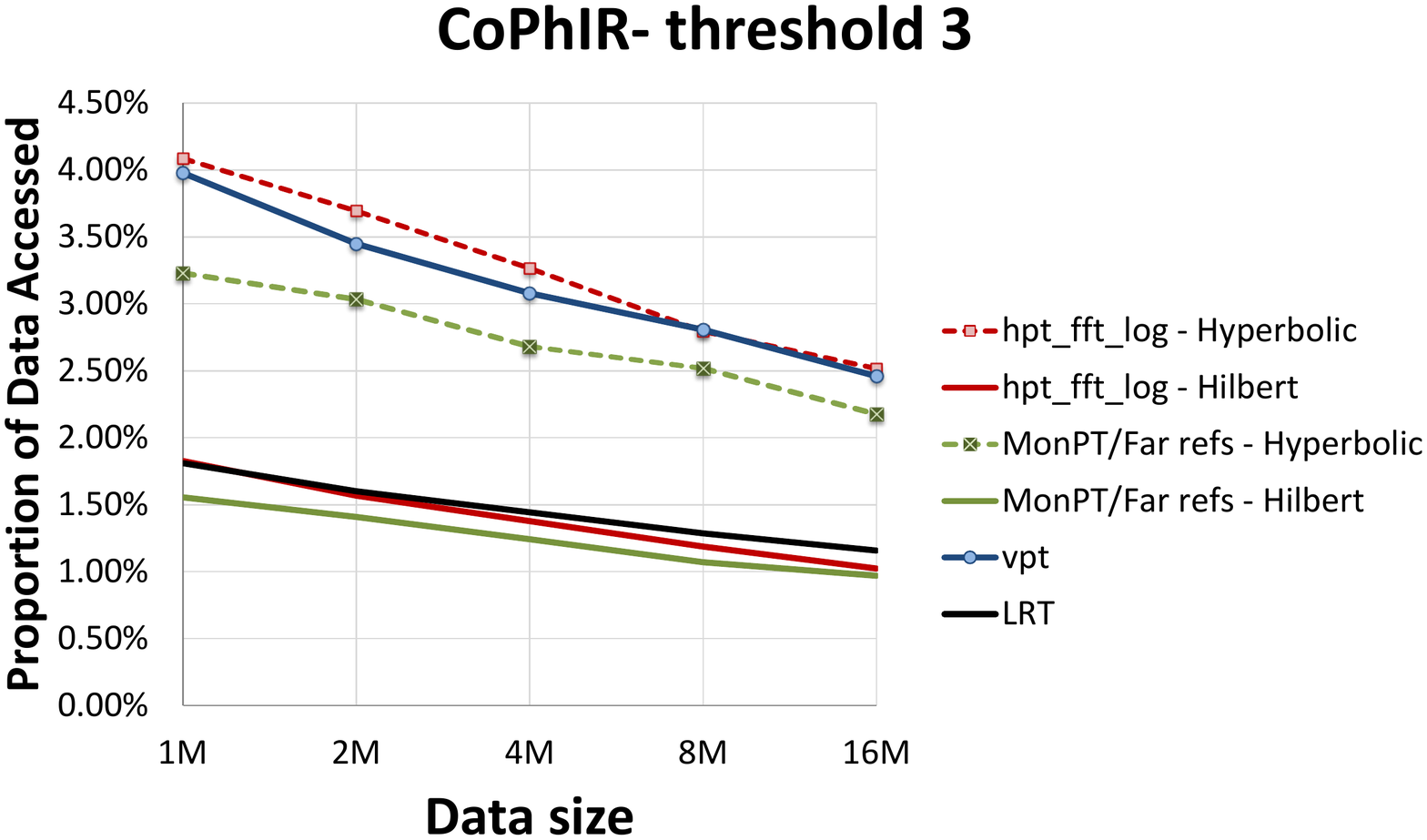}}%
		\caption{\h{Test over increasingly large subsets of CoPHIR images set. The proportion of data access per query is reported for: log-sized hyperplane trees (\texttt{hpt\_fft\_log}), monotone hyperplane trees (\texttt{MonPT/Far refs}), a vantage point tree (\texttt{vpt}) and our linear regression tree (\texttt{LRT}).}}
		\label{figure_scaling}
	\end{center}
\end{figure*}

%
%


\section{Conclusions} \label{sec:conclusion}
We have presented  a novel observation based on the four-point property that is possessed by many useful distance metrics. We have shown how the property that any four points from the original space may be embedded in \threed {} as a tetrahedron leads to further geometric guarantees, in particular we have shown a lower-bound distance that can be calculated from knowledge of the sides of two tetrahedral faces. We have shown a few examples of how metric indexes can be constructed from this property, and have achieved new best performances for Euclidean distance search over two of the SISAP benchmarks. Further we have demonstrated that the advantages shown over the relatively small and tractable benchmark sets extend to larger, less tractable spaces.

There are some new areas of investigation opened up by this work. Further study of the  use of different partition strategies used to fit the reference points and data available at each node of an indexing structure should be worthwhile. Given the supermetric properties,  much more information is available during tree construction than we have, so far, fully exploited. In particular, given an analytic expression for the discarding rule,  a term for the distance between reference points, and  various assumptions about the searching radius and the distance of the query to the reference points  it should be possible to maximise the discarding power of the node. This would allow the construction of a controlled balancing which will outperform any randomly unbalanced index structure. We have not yet investigated the possibility of controlling the balance within n-ary partition trees, or applying domain-specific partition strategies to them, which seem to be the most promising avenues for achieving still better exact search performance.

Finally, we are excited by the possibility of extending this work into higher dimensions. In  all but pathologically constructed cases, a space with the four-point property also has the so-called $n$-point property: that is, any $n + 1$ points may be isometrically embedded in $n$-dimensional Euclidean space. We are currently investigating various geometric guarantees that can be determined in arbitrarily high dimensions.

\section*{Acknowledgements}
We would like to thank the anonymous referees for helpful comments on an earlier version of this paper. \h{We are particularly grateful to Dr. Fabrizio Falchi for his help in accessing the CoPhIR data set.}
Richard Connor would like to acknowledge support by the National Research Council of Italy (CNR) for a Short-term Mobility Fellowship (STM) in June 2015, which funded a stay at ISTI-CNR in Pisa during which   this work was conceived.
The work was also partially funded by Smart News, ``Social sensing for breaking news", co-funded by the Tuscany region under the FAR-FAS 2014 program, CUP CIPE D58C15000270008.

\bibliographystyle{elsarticle-num}
\bibliography{bib/connor,bib/general}
\end{document}